\documentclass[twocolumn]{IEEEtran}
\usepackage{amsmath}
\usepackage{algorithmicx}
\usepackage{algpseudocode}

\usepackage[pdftex]{graphicx} 
\usepackage{array} 
\usepackage[caption=false,font=footnotesize]{subfig} 
\usepackage{stfloats} 
\usepackage{url} 

\usepackage{microtype}
\usepackage[utf8]{inputenc} 
\usepackage[T1]{fontenc}    
\usepackage{paralist} 
\usepackage{marvosym} 
\usepackage{bm} 
\usepackage{bbm} 
\usepackage{amssymb}   
\usepackage{mathrsfs} 
\usepackage{graphicx}
\usepackage{tabularx} 
\usepackage{booktabs} 
\usepackage[flushleft]{threeparttable} 
\usepackage{colortbl} 
\usepackage{makecell} 
\usepackage{multirow} 
\usepackage{nicefrac} 
\usepackage[dvipsnames]{xcolor} 
\usepackage{soul} 
\setul{0.5ex}{0.3ex}
\definecolor{citecolor}{RGB}{0,0,255}
\usepackage[pagebackref=false,breaklinks=true,colorlinks,
    citecolor=citecolor,bookmarks=false]{hyperref}
\usepackage{cleveref}  
\crefformat{figure}{Fig.~#2#1#3}
\crefformat{equation}{Eq.~(#2#1#3)}
\crefformat{table}{Table~#2#1#3}
\crefformat{section}{Section~#2#1#3}
\crefformat{algorithm}{Algorithm~#2#1#3}
\crefformat{appendix}{Appendix #2#1#3}
\crefformat{subsection}{subsection~#2#1#3}
\usepackage{cellspace}
\usepackage[numbers,sort&compress]{natbib} 

\usepackage{xspace} 

\usepackage{pifont} 
\newcommand{\xmark}{\ding{53}}%

\definecolor{Gray}{rgb}{0.9,0.9,0.9}
\definecolor{LightCyan}{rgb}{0.88,1,1}
\newcolumntype{a}{>{\columncolor{Gray}}c}
\newcolumntype{x}{>{\columncolor{white}}c}

\begin{document}

\setlength{\abovedisplayskip}{.5\baselineskip} 
\setlength{\belowdisplayskip}{.5\baselineskip} 

\title{SWAN: Synergistic Wavelet-Attention Network for Infrared Small Target Detection}


\author{
  Yuxin Jing,
  Jufeng Zhao,
  Tianpei Zhang,
  Yiming~Zhu
  

}

\maketitle

\section{abstract} \label{sec:abstract}
\begin{abstract}
Infrared small target detection (IRSTD) is thus critical in both civilian and military applications. This study addresses the challenge of precisely IRSTD in complex backgrounds. Recent methods focus fundamental reliance on conventional convolution operations, which primarily capture local spatial patterns and struggle to distinguish the unique frequency-domain characteristics of small targets from intricate background clutter. To overcome these limitations, we proposed the Synergistic Wavelet-Attention Network (SWAN), a novel framework designed to perceive targets from both spatial and frequency domains. SWAN leverages a Haar Wavelet Convolution (HWConv) for a deep, cross-domain fusion of the frequency energy and spatial details of small target. Furthermore, a Shifted Spatial Attention (SSA) mechanism efficiently models long-range spatial dependencies with linear computational complexity, enhancing contextual awareness. Finally, a Residual Dual-Channel Attention (RDCA) module adaptively calibrates channel-wise feature responses to suppress background interference while amplifying target-pertinent signals. Extensive experiments on benchmark datasets demonstrate that SWAN surpasses existing state-of-the-art methods, showing significant improvements in detection accuracy and robustness, particularly in complex challenging scenarios.

\end{abstract}


\section{Introduction} \label{sec:introduction}



Infrared imaging system can capture thermal radiation, offering the advantage of being unaffected by lighting changes and capable of penetrating visual obstacles like haze and smoke \cite{sobrino2016review}.
Infrared systems are extensively used in unmanned monitoring \cite{deng2016small}, maritime search and rescue \cite{wu2023mtu, hu2024smpisd}, remote sensing \cite{sun2020infrared}, and precision target detection \cite{zhu2024towards} due to their all-weather operability, robust anti-interference, and thermal sensitivity. As the core of infrared imaging system, infrared small target detection(IRSTD) plays an indefensible role in both civilian and military applications.




Although IRSTD has proven effective in various conditions, the targets in these scenarios often appear as small objects due to long distances. Therefore, some inherent challenges hinder the precision, potentially leading to missed detections, false alarms, and inaccurate localization. Specifically, IRSTD currently confronts three core challenges:
\begin{enumerate}

\item \textbf{Conflict between Ultra-Small Targets and Low Signal-to-Clutter Ratio (SCR)}: Due to long imaging distance, infrared small targets are small and typically exhibit low signal-to-clutter ratio (SCR), rendering them prone to being overwhelmed by substantial noise and background clutter.

\item \textbf{Scarcity of Intrinsic Target Features}: infrared small targets appear as inconspicuous blobs, lacking sufficiently prominent features to distinguish them from visually similar false alarms when relying solely on localized target information.

\item \textbf{Coupling of Multi-Source Interferences with Background Homogeneity}: Infrared images are often plagued by numerous interferences resembling the targets of interest, such as noise points and hot spots, making it nearly impossible to distinguish them from real targets based solely on localized visual comparison.

\end{enumerate}


In recent decades, IRSTD can be separated into two major phases: model-driven methods and data-driven methods. Model-driven methods can be further categorized into three sub-classes: filtering-based methods, local information enhancement, and low-rank sparse decomposition. Filtering methods\cite{zhu2020balanced, lu2022enhanced, deng2021entropy, deng2021infrared, bai2010analysis, li2021infrared, zhang2023infrared, ren2020infrared} utilize morphological operations or special designed filter to suppress the background and enhance the target feature. Meanwhile, local information enhancement methods  \cite{chen2013local, wei2016multiscale, qiu2022global, qiu2020adaptive, xu2023infrared, lu2023infrared} generate saliency maps by calculating contrast values within local windows, thereby amplifying the distinction between the target and the background. Furthermore, low-rank decomposition methods \cite{dai2017reweighted, zhang2018infrared, zhang2019infrared, zhong2023infrared} model the image as a linear combination of a low-rank background matrix and a sparse target matrix, separating the target and background components through optimization algorithms. Although model-driven methods can effective in specific scenarios, their performance relies heavily on manually designed feature priors. In complex and dynamic scenes, they face core bottlenecks such as high sensitivity to parameters and weak generalization ability.

\begin{figure}
    \centering
    \includegraphics[width=1\linewidth]{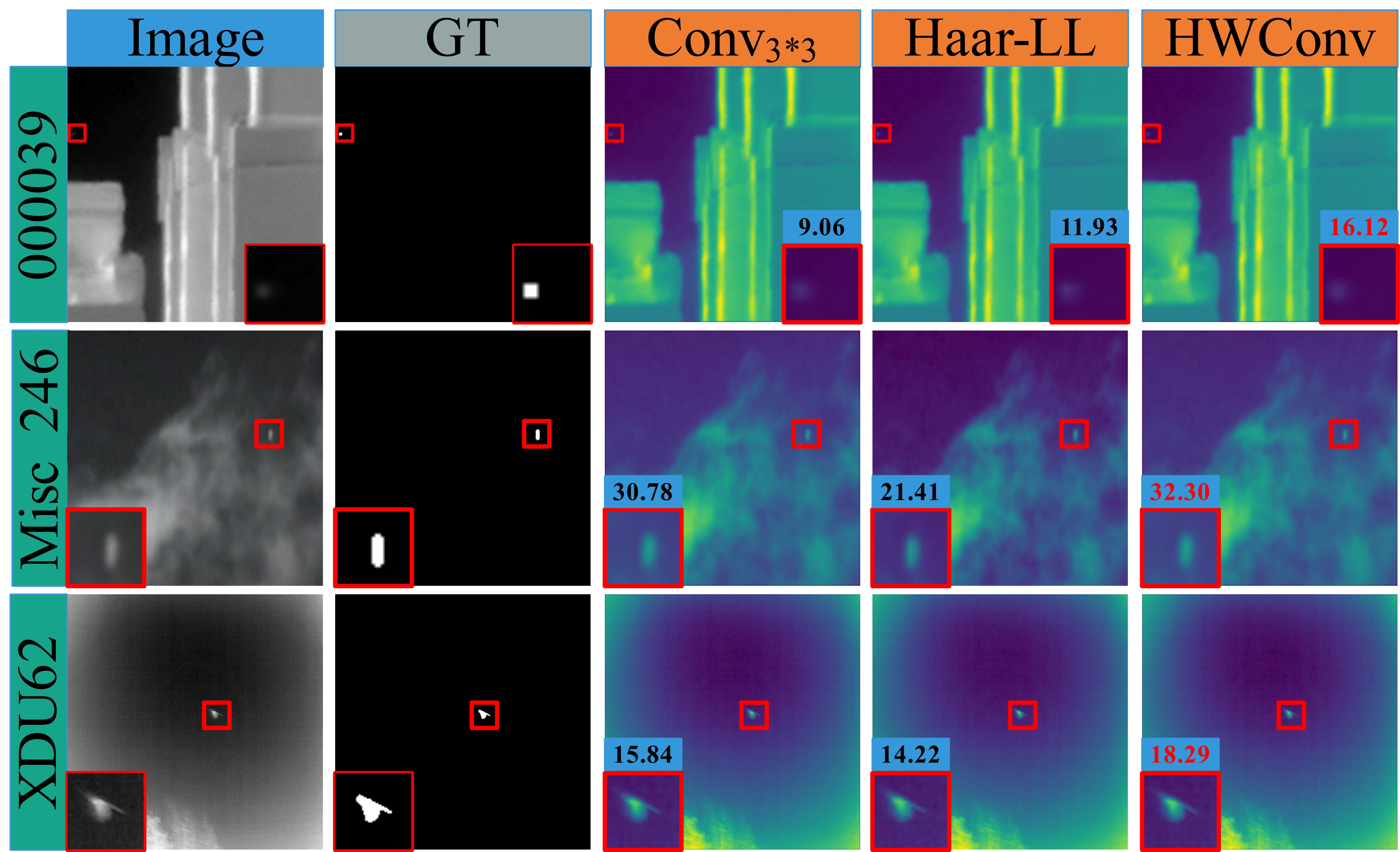}
    \caption{Heat map example. The columns from left to right
respectively represent the original image, the ground truth, the
heat map output by conventional convolution, the heat map
output by the low-frequency subband of Haar wavelet, and the
heat map output by HWConv. Zoomed-in regions of interest
are annotated with their corresponding localized Signal-
to-Clutter Ratio(SCR) values.}
    \label{fig2}
\end{figure}

Driven by the flourishing development of deep learning, data-driven methods have demonstrated significant progress in IRSTD, leveraging strong generalization capabilities and efficiency. Overall, state-of-the-art approaches predominantly focus on exploiting local spatial features. This includes enhancing the response to high-frequency edge and shape information \cite{li2022dense, zhang2022isnet, zhu2024towards}, designing modules to amplify local contrast \cite{dai2021attentional, zhang2023attention, bai2022cross}, and leveraging feature gradients \cite{bai2022cross, li2022dense}. Representative methods like DNANet \cite{li2022dense} utilize dense-connected dynamic convolution to intensify local edge features, while UIUNet \cite{wu2022uiu} employs dense skip connections in U-Net \cite{ronneberger2015u} to mitigate detail loss inherent in deep architectures. Concurrently, emerging efforts aim to address the limitation of local processing. The DATransNet \cite{hu2025datransnet} incorporates Transformer modules into CNN backbones to model long-range pixel dependencies via self-attention mechanisms. Despite these advancements, fundamental challenges largely stemming from the inherent limitations of convolution operations. Firstly, the constrained receptive fields in shallow network layers hinder their ability to capture broader contextual information and the subtle signatures of small targets effectively, leading to insufficient modeling of global contextual dependencies \cite{zhong2022detecting}. Secondly, although deeper layers achieve larger receptive fields through downsampling, this process introduces detrimental side effects: high-frequency background noise is propagated and amplified across layers \cite{liu2023infrared}, and the fine details of small targets are inevitably degraded due to repeated spatial downsampling.

Specifically, current data-driven IRSTD methods face critical limitations in feature modeling:
\begin{enumerate}

\item \textbf{Superficial Frequency-Spatial Fusion}: Current methods integrate wavelet transforms and convolutions primarily through simple concatenation or cascading \cite{zhang2025exploring}, lacking deep cross-domain feature coupling. Crucially, low-frequency subband semantics remain inadequately enhanced.

\item \textbf{Inefficient Spatial Interaction}: Modeling long-range dependencies (e.g. via global self-attention such as DATransNet \cite{hu2025datransnet} and SCTransNet \cite{yuan2024sctransnet}) incurs prohibitive computational cost, especially for high-resolution infrared images, while local convolutions inherently lack sufficient receptive fields.

\item \textbf{Neglected Channel Semantics}: Existing methods fail to account for the distinct semantics of channels (e.g. shallow high-frequency detail vs. deep low-frequency context) during multi-scale fusion. Direct fusion operations, such as addition or concatenation, suppress target-related channels in favor of dominant background channels.

\end{enumerate}

Motivated by the identified limitations in feature modeling and information interaction, we propose a novel framework designed to fundamentally enhance feature representation and integration. Our core objectives are threefold: Firstly, achieve synergistic fusion of spatial and frequency domain features, moving beyond simple concatenation to establish an intrinsic coupling mechanism. This mechanism leverages and enhances complementary information for more discriminative feature extraction. Meanwhile, enable effective modeling of long-range contextual dependencies in high-resolution images, overcoming the computational bottlenecks of global attention while surpassing the limited receptive fields of local convolutions. Furthermore, ensure semantic-aware multi-scale feature fusion, where features from different depths and semantics (e.g. fine-grained details vs. broader context) are dynamically weighted to prioritize target-relevant information and suppress dominant background interference. Collectively, these optimizations enhance the entire feature encoding-decoding pipeline, fostering robust feature representation, comprehensive contextual understanding, and semantically meaningful information aggregation, which leads to significant improvements of IRSTD.

\begin{figure}
    \centering
    \includegraphics[width=1\linewidth]{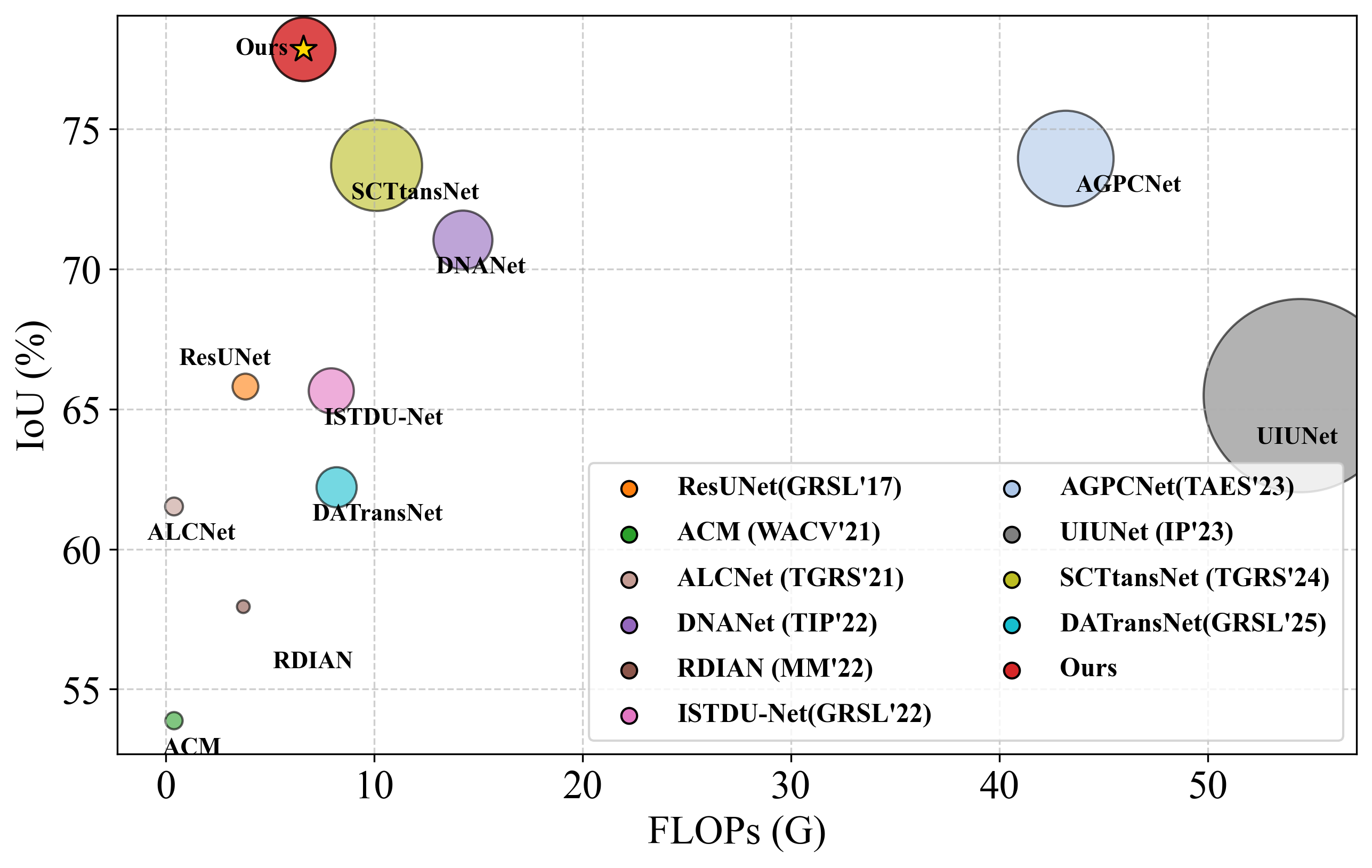}
    \caption{Comparison of IoU, parameter count (Params.), and Flops of mainstream ISTD deep learning methods on the REAL dataset.}
    \label{fig1}
\end{figure}
Based on the preceding analysis, we proposes the \textbf{S}ynergistic \textbf{W}avelet-\textbf{A}ttention \textbf{N}etwork (\textbf{SWAN}), establishing an efficient detection framework through three core innovations: At the feature extraction layer, we design the \textbf{Haar Wave Convolution (HWConv)}. This module integrates wavelet decomposition’s low-frequency subband (LL) with spatial convolution features via a cross-domain alignment unit, enabling deep coupling of low-frequency energy distribution and high-frequency details. (Visual comparison of enhanced SCR is shown in Fig. \ref{fig2}). Meanwhile, We propose the \textbf{Shifted Spatial Attention (SSA)} mechanism. SSA employs a dynamic window shifting strategy to break fixed window boundaries, achieving linear-complexity modeling of long-range dependencies across windows. This balances efficient processing of high-resolution images with contextual awareness. Moreover, We devise the \textbf{Residual Dual-Channel Attention (RDCA)} module. RDCA dynamically allocates channel-level weights to shallow (high-freq) and deep (low-freq) features using a residual path. This suppresses background-dominant channels and enhances target-relevant frequency bands for adaptive multi-scale fusion. Through collaborative optimization of innovations, SWAN significantly enhances detection under low SNR and complex backgrounds. Experimental in Sec. \ref{sec:experiment} validate the comprehensive metric comparison shown in Fig. \ref{fig1}, confirms our superior accuracy-efficiency trade-off of SWAN.

The contribution of this paper are listed as follows:

\begin{enumerate}
\item We propose HWConv, \textbf{a novel frequency-spatial fusion paradigm} that fundamentally bridges spectral energy distribution via wavelet low-frequency priors and spatial feature representation.

\item Featuring dynamically shiftable receptive fields, the proposed SSA enables \textbf{linear-complexity global context modeling} in high-resolution infrared systems. This spatial interaction mechanism fundamentally transcends the fixed-window limitations of vision Transformers.

\item Addressing channel semantic conflicts in frequency fusion, RDCA introduces a novel residual calibration framework. This \textbf{multi-scale frequency-aware architecture} suppresses background-dominated bands while enhancing target features, overcoming critical limitations in weak-signal recovery systems.

\end{enumerate}

\section{Related Work} \label{sec:related}

\subsection{Infrared Small Target Detection} \label{subsec:IRSTD}
\subsubsection{\textbf{Model-driven Methods for IRSTD}}
\label{subsubsec:model-driven}

Traditional model-driven infrard small target detection 
(IRSTD) methods primarily rely on manual feature design, whcih include filter, local information enhance, and low-rank sparse decomposition methods. For instance, filter-based methods, such as New Top-Hat filtering \cite{deng2021entropy, deng2021infrared}, Max/Mean fliter \cite{li2021infrared}, and frequency filter \cite{ren2020infrared} utilize morphological operations to separate targets from backgrounds. They match target shapes using predefined structural elements and extract residual signals through differencing. However, their performance heavily depends on the shape and size of these elements, and the differencing operation can amplify high-frequency noise, often resulting in the disappearance of target signals in low signal-to-noise ratio (SNR) scenarios. Meanwhile, local information enhancement methods, including Local Contrast Measurement (LCM) \cite{chen2013local} and multi-scale methods (RLCM) \cite{han2018infrared}, MPCM \cite{wei2016multiscale} and FLCM \cite{lu2023infrared} generate local saliency maps by calculating contrast within local windows, thereby amplifying local contrast responses in target areas. Yet, they are sensitive to complex backgrounds, like cloud textures, and struggle to adapt to variations in target scale. Moreover, Low-rank sparse decomposition methods such as IPI \cite{gao2013infrared} and RIPT \cite{dai2017reweighted} model, aim to split images into low-rank background and sparse target components. However, small targets, which have minimal area and weak grayscale intensity, weak target energy often leaks into the background component during decomposition, leading to missed detections. Overall, the model-driven methods is sensitive to interfere of backgrounds and lack adaptive parameter tuning to change of different conditions.

\subsubsection{\textbf{Deep Learning Methods for IRSTD}}

With the improvement of deep learning technology, existing data-driven methods have achieved significant progress.
The data-driven methods make special designed improvement based on the feature of small targets.
For instance, the Asymmetric Context Modulation (ACM) network \cite{dai2021asymmetric} introduces an asymmetric feature fusion technique, an alternative to conventional skip connections in U-Net.
Meanwhile, Attention Local Contrast (ALCNet) \cite{bai2010analysis} utilizes fixed-weight multi-scale feature concatenation without dynamically adjusting.
Moreover, Unet-in-Unet (UIUNet) \cite{wu2022uiu} enhances the detection of small target contrasts by integrating multiple U-Net structures and using interactive cross-attention mechanisms for feature fusion.
Furthermore, Gated-Shaped Trans-Net (GSTUnet) \cite{zhu2024towards} merges vision Transformer technology with CNNs in the encoder to efficiently extract edge features from small targets. 
Additionally, Dynamic Attention Trans-Net (DATransNet) \cite{hu2025datransnet} incorporates Transformer modules into CNN backbones to model long-range pixel dependencies via self-attention mechanisms.
Current data-driven IRSTD methods improve target detection robustness through multi-scale feature fusion. 


In conclusion, the improvement of IRSTD by data-driven methods focus on edge enhancement \cite{zhang2022isnet,zhu2024towards, xiao2024background}, local information enhancement \cite{dai2021attentional, zhang2023attention, bai2022cross, zhu2021dau, hu2025datransnet}, deeper network structure \cite{li2022dense, wu2022uiu}. However, constrained by the local perceptual nature of convolutional operations, they struggle to model long-range contextual dependencies. In complex backgrounds, high-frequency noise regions may be misclassified as targets, leading to false alarms. Our method is distinguished from other methods from following key components. First, Stronger feature extraction while preserving small target features in the deeper layers of the network. Second, dynamic feature selection to accurate differentiate between small target features and false positives.

\subsection{Wavelet Transform with Deep Learning}

The Discrete Wavelet Transform (DWT) decomposes infrared images into multi-frequency subbands (LL, LH, HL, HH), enabling simultaneous capture of low-frequency radiation trends and high-frequency edge/texture details \cite{zhang2025wife}. Current integration paradigms demonstrate two primary approaches: Early-stage decomposition methods like WaveCNet \cite{yang2023wavecnns} feed decomposed subbands into parallel convolutional branches, preserving initial frequency characteristics but with limited coordination to deep semantic features. Alternatively, attention-guided frameworks such as WA-CNN \cite{zhao2022wavelet} utilize wavelet coefficients to weight spatial attention maps, enhancing target region focus while underutilizing hierarchical subband dependencies. These implementations reveal untapped opportunities for deeper synergy, particularly in establishing dynamic cross-band interactions throughout network depths, jointly optimizing wavelet and convolutional representations, and exploiting multi-scale correlations inherent in subband relationships for IRSTD.

Distinctively, our framework pioneers a deep frequency-spatial coupling mechanism. we design a wavelet-convolution parallel nesting module that jointly encodes the LL subband and spatial features via cross-domain alignment. This resolves the semantic disconnection in data-driven methods by: \textbf{Dynamic interaction}: Efficient feature recalibration between energy distribution (LL) and structural details. \textbf{Hierarchical coordination}: Multi-scale dependency propagation across network depths. \textbf{Saliency enhancement}: Significant SCR elevation (Fig. \ref{fig2}) through synergistic target amplification.

\subsection{Vision Transformer in IRSTD}


While convolution-based attention Sequeeze-Extraction 
(SE) \cite{hu2018squeeze} or Convolution Block Attention Module(CBAM) \cite{woo2018cbam}) suffers from fixed receptive fields and channel bias, recent Transformer \cite{vaswani2017attention} variants address these limitations through distinct strategies. For instance, the SCTransNet \cite{yuan2024sctransnet} establishes spatial-channel interactions via cross-attention but requires multi-stage processing for global context. Meanwhile, the DATransNet \cite{hu2025datransnet} Combining CNN to extract local details and Transformer to model global dependencies. The Swin Transformer \cite{liu2021swin} adopts cross window sliding self attention computation, which can perform global feature extraction with only linear computational complexity.However, critical limitations still exist, such as the complexity accuracy balance of STCransNet and the rigid window constraints that limit dynamic target adaptation.

The key difference between our method and others lies in the SSA module, which breaks the limitations of traditional window-based self-attention. Meanwhile, compared to Swin Transformer, SSA fundamentally advances contextual modeling beyond it by enabling single-layer cross-window integration through adaptive cyclic shifting. This architectural divergence yields critical advantages IRSTD.


\section{Method} \label{sec:method}

\subsection{Overall Structure of SWAN}
The overall structure of SWAN consists of three main components: Haar Wavelet Convolution (HWConv) for feature extraction, Shifted Window Self-Attention (SSA) for spatial dependency modeling, and Residual Dual-Channel Attention (RDCA) for channel semantic calibration. HWConv integrates principles from both spatial and frequency domains, focusing on the precise identification of infrared small target features. SSA establishes long-range dependencies between features of different scales during the encoding-decoding phase, overcoming the interference signals introduced by skip connections in traditional UNet structures. RDCA serves as the final fusion module, employing a dual residual structure to compute channel attention weights for both upsampled features and skip connection features, facilitating the effective fusion of shallow and deep features.

\begin{figure*}
    \centering
    \includegraphics[width=1.0\linewidth]{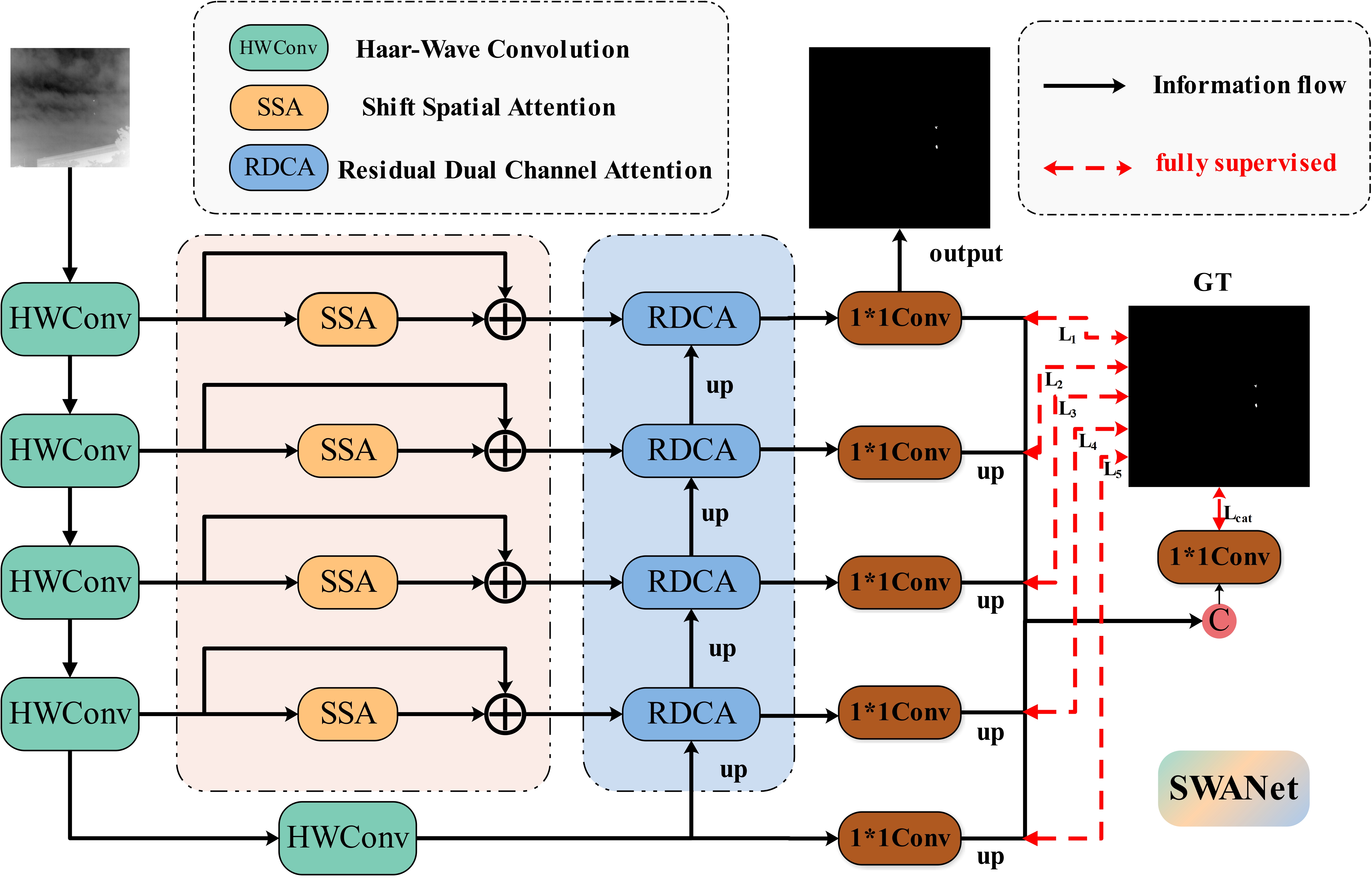}
    \caption{The overall architecture of the proposed SWAN features a UNet framework with skip connections, which includes the HWConv, SSA, RDCA modules, as well as a fully supervised loss function.}
    \label{overall}
\end{figure*}

As shown in Fig. \ref{overall}, the input image undergoes a two-level nested fusion with HWConv, except the first iteration utilizing the low-frequency subbands from wavelet transforms for deep feature extraction. At the skip connection points, SSA is introduced to enhance target representation while using residual connections to align the enhanced target features. Subsequently, RDCA is applied to calculate channel attention weights, thereby improving the fusion of deep and shallow features. During the training phase, a $1 \times 1$ convolution operation is performed on the output features of each layer, and these are concatenated to compute the loss using Binary Cross-Entropy (BCE) for deep supervision to optimize SWAN.

\subsection{Module-wise Introduction In SWAN}

\subsubsection{\textbf{HWConv}}
Existing methods for fusing wavelet subbands with convolutional features often fail to fully exploit low-frequency energy distributions and high-frequency details of targets due to inadequate cross-domain semantic interaction caused by simplistic concatenation or single-domain processing. To address this, we propose the HWConv module featuring an innovative parallel nested architecture that integrates wavelet decomposition with spatial convolution. By incorporating a cross-domain feature alignment unit, our approach enforces deep coupling between modalities, effectively bridging the semantic gap between frequency-domain energy patterns and spatial structural information.

Considering the preference of Convolutional Neural Networks (CNNs) for high-frequency features, while infrared imaging small target characteristics tend to favor low-frequency features \cite{finder2024wavelet}, HWConv enhances the ability to capture small target features by introducing Haar wavelet transforms. This method employs a parallel nested approach combining Haar wavelets and convolution, significantly improving feature detection. The integration of frequency domain analysis with convolution facilitates broader feature capture through cross-information channel fusion, which notably boosts the efficiency of feature extraction. Moreover, as the computation load of convolution increases with larger kernel sizes, utilizing low-frequency subband features not only allows for effective feature extraction but also reduces computational complexity.
The formula for the computational complexity (Flops) of standard convolution is:
\begin{equation}
    N_{\text{flops}} = C_{in} \times C_{out} \times K^2 \times H \times W
    \label{eq:flops}
\end{equation}
where $C_{in}$ represents the number of input channels, $C_{out}$ represents the number of output channels, 
$K^2$ denotes the size of the convolution kernel, and H and W are the height and width of the feature map, respectively.
On the contrast, the formula for the parameter count of HWConv is:
\begin{equation}
   4C \cdot \sum_{l=0}^{l=i} \frac{H}{2^l} \cdot \frac{W}{2^l}
\end{equation}
where C represents the number of channels for wavelet transforms at each layer. Each wavelet transform produces four subbands. $l$ denotes the number of nested levels, while H and W are the height and width of the feature map, respectively.

\begin{figure}
    \centering
    \includegraphics[width=1\linewidth]{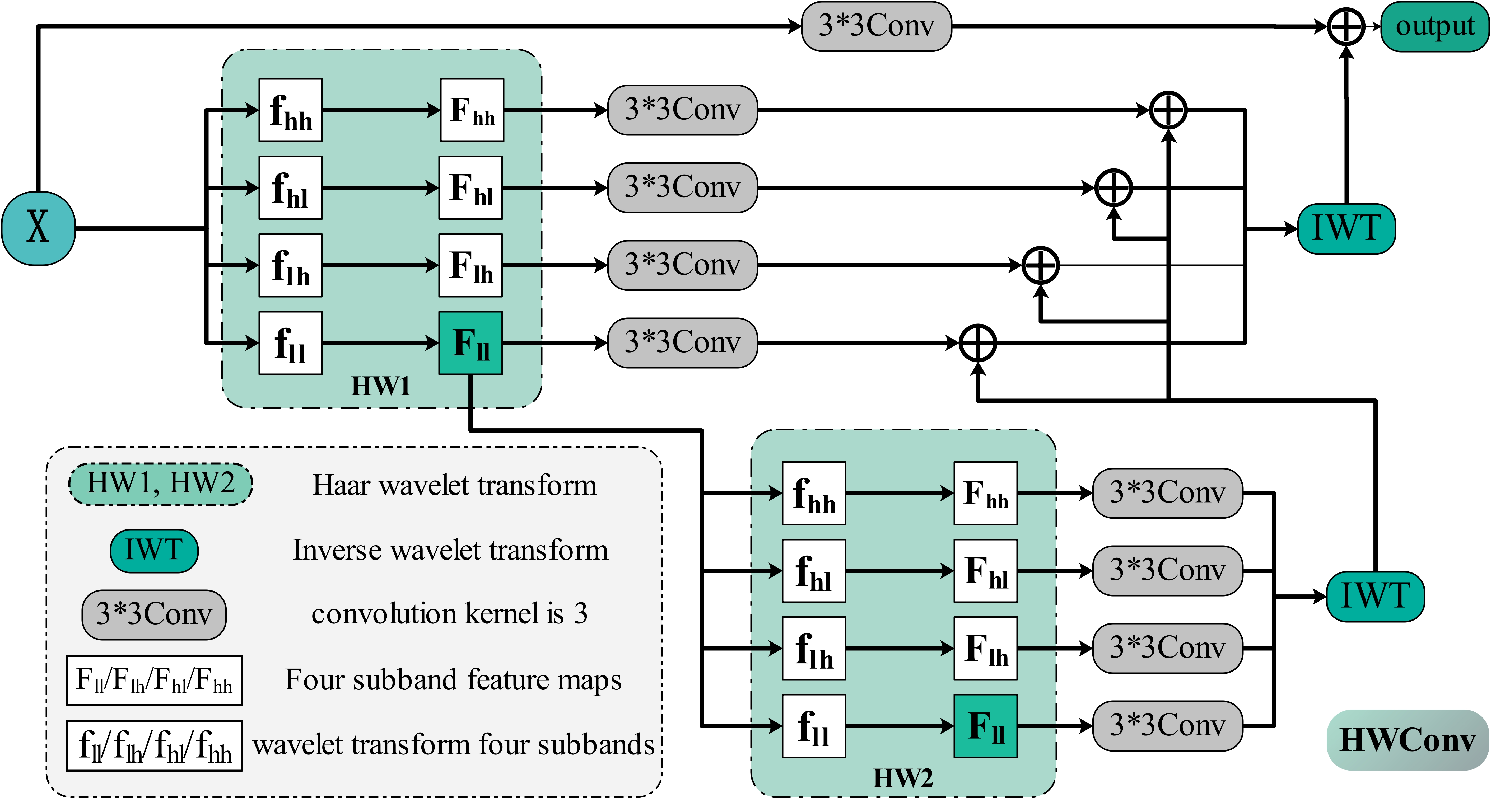}
    \caption{The HWConv module structure consists of a two-level nested mode combining frequency domain wavelet transformation and convolution.}
    \label{HWConv}
\end{figure}

The specific steps are illustrated with a two-level nesting example (as shown in the Fig. \ref{HWConv}). First, the input infrared image undergoes a Haar wavelet transform (HW1), resulting in four subbands. The low-frequency subband captures the overall trend of the signal, allowing for further information extraction. The remaining high-frequency subbands are processed through convolution and then used for inverse transformation to reconstruct features.

\textbf{Step 1}: Perform Haar wavelet transformation on the input image.
\begin{equation}
X^{C*H*W} \xrightarrow{\text{Haar}} \left\{ F_{1}^{LL}, F_{1}^{LH}, F_{1}^{HL}, F_{1}^{HH} \right\} \\
\end{equation}

The initial input features X are transformed using the Haar wavelet transform to obtain the high and low frequency semantic features $F_{1}^{LL}$, $F_{1}^{LH}$, $F_{1}^{HL}$, $F_{1}^{HH}$.

\textbf{Step 2}: The nested wavelet transformation uses the previous level's low-frequency subband. The second wavelet transform (HW2) focuses on decoding the LL1 low-frequency subband to further locate targets within the image. 

\begin{equation}
F_{k}^{LL} \xrightarrow{\text{Haar}} \left\{ F_{k+1}^{LL}, F_{k+1}^{LH}, F_{k+1}^{HL}, F_{k+1}^{HH} \right\} \\
\end{equation}

Any layer (the k-th layer) $F_{k}^{LL}$ of the low-frequency subband undergoes wavelet transformation to output the frequency domain features of the (k+1)-th layer.

\textbf{Step 3}: Simultaneously, the high and low-frequency bands obtained from the two HW transforms undergo classic 3×3 convolution operations for local feature extraction. 
\begin{equation}
 F_k = \text{Conv}_{3 \times 3} \left( F_k^{LL}, F_k^{LH}, F_k^{HL}, F_k^{HH} \right);\\
\end{equation}

Finally, each convolved image group will undergo inverse wavelet transformation and be fused with the features from the upper layers through addition, resulting in the final output.

\begin{footnotesize}
\begin{equation}
    \text{output} = F_1 \oplus IWT(F2) \cdots \\IWT(F_{k} \oplus IWT(F_{k+1})) \cdots)
\end{equation}
\end{footnotesize}

The final output result is obtained by progressively adding the reconstructed features from the deepest layer, which have undergone $3 \times 3$ convolution, to the previous layer. Here, $F_{k}$ represents the features processed by the $3 \times 3$ convolution, and IWT denotes





\subsubsection{\textbf{SSA}}
Although HWConv integrates frequency-spatial information, the weak responses of small targets within its composite feature maps urgently require efficient long-range modeling to enhance discriminability under complex backgrounds. To bridge this gap, motivate by swin-transformer \cite{liu2021swin}, we proposed SSA mechanism employs a dynamic window shifting strategy that seamlessly interfaces with the HWConv feature flow. This enables precise aggregation of scattered target cues while suppressing background interference, significantly augmenting contextual awareness of targets.

\begin{figure}
    \centering
    \includegraphics[width=1\linewidth]{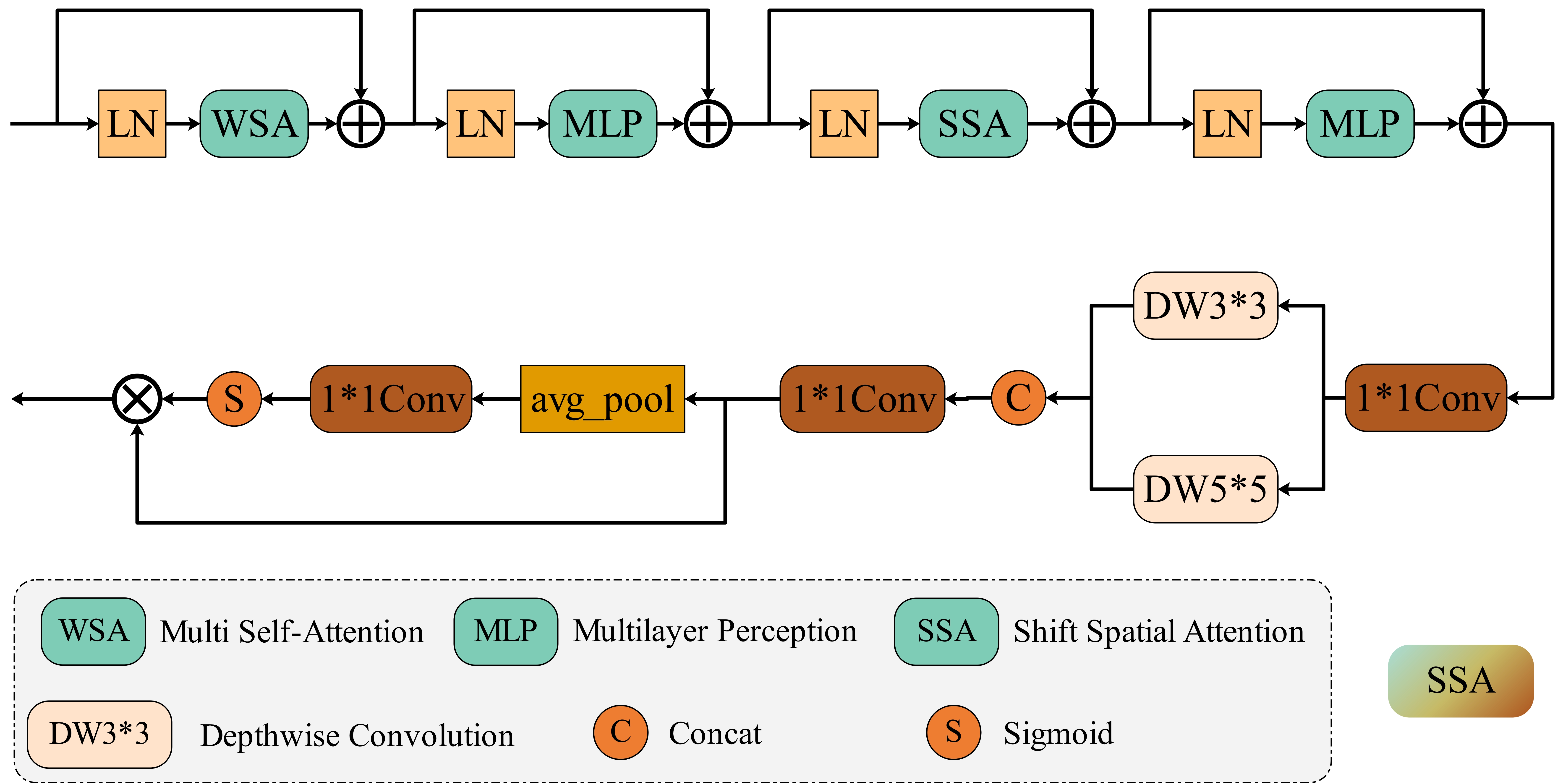}
    \caption{The Shifted Spatial Attention module structure features a dual attention mechanism that facilitates spatial information interaction and the integration of channel information.}
    \label{fig:SSA}
\end{figure}

From the holistic information flow perspective of the SSA mechanism, the Window Self-Attention (WSA) operates by partitioning the input image into windows and computing self-attention within each window. To enhance cross-window information interaction, the SSA module employs a dynamic window shifting strategy. This allows the computed attention to span multiple windows, thereby strengthening spatial information propagation across the feature map. Following the computation of both self-attention and cross-attention operations, the feature map undergoes channel-wise averaging via a $1 \times 1$ convolution. It then proceeds through parallel $3 \times 3$ and $5 \times 5$ depthwise separable convolution (DWConv) operations. Residual connections preserve potentially lost features, while average pooling eliminates redundant information, ultimately yielding the optimized feature output.
The overall model is shown in the Fig. \ref{fig:SSA}.

WSA performs the calculation of the self-attention mechanism by dividing fixed-size patch blocks. The query vector (Q), key vector (K), and value vector (V) are obtained through linear transformation of the input matrix X. Based on the query vector Q, the similarity between the query vector and all key vectors K is calculated to obtain a weight distribution, which is used to weighted sum the associated value vectors V. The formula for WSA to calculate self-attention is as follows:
\begin{equation}
\text{Atten}_{\text{WSA}} = \text{softmax}\left( \frac{Q \cdot K^T}{\sqrt{d_k}} \right) V \\
\end{equation}
where the softmax function normalizes the Query-Key similarity scores to generate an attention weight distribution. The scaling factor $\sqrt{d_k}$ is used to prevent gradient vanishing caused by excessively large dot product values.

Based on WSA, we found that it ignored the information interaction between each patch block. Therefore, the SSA shifted window attention mechanism was added to enhance the spatial information interaction. The formula for calculating the interaction attention of SSA is:
\begin{equation}
\text{Atten}_{\text{SSA}} = \text{softmax}\left( \frac{Q \cdot K^T}{\sqrt{d_k}} + D \right) V\\
\end{equation}
where the bias term D is a learnable parameter matrix. Its core function is to augment the attention weight matrix ${Q \cdot K^T}$ with a spatial-relative-position bias, rather than performing absolute positional encoding. Specifically, it encodes the relative positional relationship between any two pixels (tokens) within the same window. For an $M \times M$window containing \(M^{2}\) pixels: when computing the attention weight between the i-th pixel and j-th pixel, their relative displacement $(\Delta x, \Delta y)$ in the window grid is calculated. For instance, if pixel j is positioned 1 unit right and 2 units below pixel i, then $((\Delta x, \Delta y) = (1, 2)$. The element $D[\Delta x, \Delta y]$ corresponding to this displacement is then added to the (i, j)-th entry of the ${Q \cdot K^T}$ matrix.

Overall formula of the final module:
\begin{footnotesize}
\begin{equation}
    F_S^k =MLP\left(SSA\left(MLP\left(WSA(F_H)\right)\right)\right),    k=1,2...5
\end{equation}
\end{footnotesize}
where $F_H$ represents the features after undergoing HWConv processing, while WSA and SSA denote Window Self-Attention and Shifted Spatial Attention, respectively. MLP refers to a Multi-Layer Perceptron.

\begin{footnotesize}
\begin{equation}
    F_S^{kk} = Conv_{1\times1} \left(Cat \left(DW \left( chunk(Conv_{1\times1} \left(F_S^k \right)\right) \right)\right)
\end{equation}
\end{footnotesize}

Perform a 1×1 convolution operation $Conv_{1\times1}$ on $F_S^{kk}$, split it into two parts along the channel dimension using chunk, then perform a depth-wise convolution (DW) followed by concatenation (Concat).

\begin{footnotesize}
\begin{equation}
    F_k^S =  F_S^{kk}\otimes Sigmoid(Conv_{1\times1}(avgpooling(F_S^{kk})))
\end{equation}
\end{footnotesize}
where $F_k^{S}$represents the final output result after SSA, $F_S^{k}$ and $F_S^{kk}$ denotes the features undergoing processing through each layer of SSA. After undergoing average pooling and applying the Sigmoid activation, a residual connection is used to perform element-wise multiplication $\otimes$ with the features before pooling. 

\subsubsection{\textbf{RDCA}}
\begin{figure}
    \centering
    \includegraphics[width=1\linewidth]{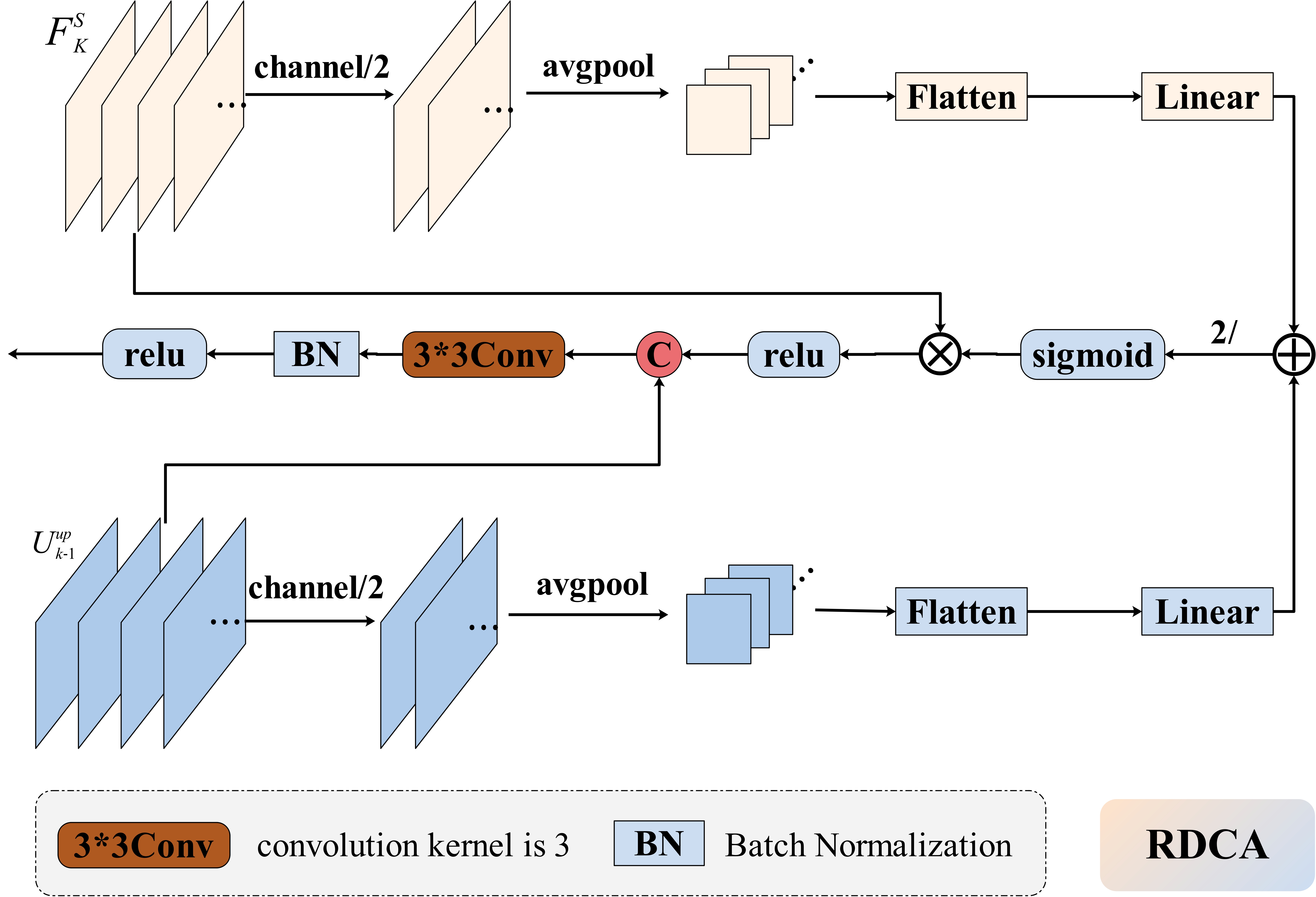}
    \caption{The Residual Dual-Channel Attention module uses adaptive weights to fuse deep and shallow feature information.}
    \label{fig:RDCA}
\end{figure}
The proposed RDCA module employs a dual-path independent calibration structure: one path enhances shallow high-frequency edge details, while the other suppresses redundant background channels in the deep features enhanced by SSA. Through residual connections, RDCA preserves critical information while optimally fusing HWConv's original features with SSA enhanced contextual features. This outputs highly discriminative target representations, supplying the detection head with robust decision-making cues.


The design of the Residual Dual-Channel Attention (RDCA) module aims to effectively fuse deep and shallow features. Deep features are combined with shallow features through upsampling to enhance feature representation. RDCA primarily generates channel-level weights by calculating a channel attention mechanism, allowing for weighted adjustments of the input features. 

The specific steps of the model are shown in the Fig. \ref{fig:RDCA} below: include halving the channels of the input tensor and applying average pooling to compress the spatial dimensions to 1. A flatten and linear processes the pooled features to extract channel-level characteristics. 

\begin{equation}
 F_k^{DD}=Linear\left(Flatten\left( avgpool\left( U_{k-1}^{up} \right)^{c/2 \times H \times W} \right)\right)
 \\
\end{equation}

\begin{equation}
 F_k^{SS}=Linear\left(Flatten\left( avgpool\left( F_{k}^{S} \right)^{c/2 \times H \times W} \right)\right)
 \\
\end{equation}
where $U_{k-1}^{\text{up}}$represents the upsampling of deep features, $F_{k}^{\text{S}}$ denotes the shallow features after the SSA module, avgpool is average pooling.

Shallow and deep features channel are then averaged to achieve channel feature fusion. The fused features are mapped to the range [0, 1] using a sigmoid function to generate channel attention weights (scale), which are then applied to adjust the shallow tensor based on these weights. Finally, a relu activation function is used to perform a nonlinear transformation on the weighted features, further enhancing their expressive power. 
\begin{equation}
F_k^{\text{DS}}=F_k^{\text{DD}} \oplus F_k^{\text{SS}}\\
\end{equation}

\begin{equation}
F_{k}^{\text{RR}}=\text{relu}\left( \text{sigmoid}\left( F_{k}^{\text{DS}} \right)^{c/2 \times H \times W} \otimes F_{k}^{\text{S}} \right)\\
\end{equation}
where $F_{k}^{\text{DS}}$ represents the processing of adding shallow $F_k^{\text{SS}}$ and deep $F_k^{\text{DD}}$ features at the channel level, sigmoid maps any real number to the interval (0,1), and rule refers to the activation function.

\begin{equation}
 F_k^R=relu\left(BN\left(\text{Conv}_{3 \times 3}\left( \text{Concat}\left( U_{k-1}^{\text{up}}, F_{k}^{RR} \right) \right)\right)\right)
\end{equation}
where $F_{k}^{\text{RR}}$ represents the result after RDCA weight allocation, $U_{K-1}^{\text{up}}$ denotes the upsampling of deep features, C represents the concatenation of shallow and deep features, and $Conv_{3\times3}$ refers to the $3\times3$ convolution followed by batch normalization (BN) and relu activation to produce the final output.

Through the collaborative optimization of frequency domain decoupling, dynamic interaction, and channel calibration, this framework significantly improves target detection accuracy and robustness in low signal-to-noise ratio and complex background scenarios.




\subsection{Loss function}
To enhance model accuracy, we adopt a fully supervised learning paradigm with the binary cross-entropy (BCE) loss as the baseline objective. BCE, a loss function widely used for binary classification tasks, fundamentally quantifies the discrepancy between model predictions and ground-truth labels. Building upon this foundation, we further leverage multi-level features for loss computation and gradient updates to guide model parameter optimization. The BCE loss is formally defined as follows:

\begin{equation}
  L_{BCE} = -(y \log(p(x)) + (1-y) \log(1-p(x)))
\end{equation}
where y is the ground truth of the sample, when the ground truth y of a sample is 1, the loss simplifies to $-y \log(p(x))$, As the predicted probability $p(x)$ approaches 1, the loss decreases; as $p(x)$ approaches 0, the loss increases sharply. Conversely, when the sample's label is y=0,  the loss reduces to $-log(1-p(x))$. Here, the loss decreases when 
$p(x)$ approaches 0, while it increases dramatically when $p(x)$ approaches 1.

\begin{equation}
  L_{k} = L_{BCE}(F^{R}_{k}, Y_{gt}), \; k = 1, 2 \ldots 5
\end{equation}
\begin{equation}
  L_{cat} = L_{BCE} \left[ \text{Sigmoid}(\text{Conv}_{1 \times 1}(F^{R}_{1}, F^{R}_{2} \ldots F^{R}_{5})) \right]
\end{equation}
\begin{equation}
  L_{total} = \sum L_{k} + L_{cat}
\end{equation}

where $L_{k}$ represents the feature loss at each layer of the model. $L_{cat}$ denotes the collective loss from concatenated feature maps across five layers. $F_{k}^{R}$ indicates the detection loss after RDCA processing at each layer's output. $Y_{gt}$ is the ground truth feature mask. The total loss is defined as the sum of individual layer losses and the concatenated feature loss.


\begin{table*}[b]
  \setlength{\abovecaptionskip}{0cm}  
  \renewcommand\arraystretch{1.2}
  \footnotesize
  \centering
  \vspace{-1\baselineskip}
\caption{Configuration For All Comparative Experiments}
\vspace{-1\baselineskip}
\label{tab:parameters}
\tabcolsep=0.3cm
\renewcommand\arraystretch{1.3}
\begin{center}
\begin{tabular}{ll}
\toprule[1pt]
Methods  & Key parameters configurations   \\ \hline
\multicolumn{2}{l}{\textit{Model-Driven methods}}  \\ \hline 
FKRW \cite{qin2019infrared}(2019) & Windows size: 11, $K=4,p=6,\beta=200$ \\ 
IPI \cite{gao2013infrared}(2013) & Patch size: 50, sliding step: 10, $\lambda=1/\sqrt{\mathrm{min}(m,n)}$, $\epsilon=10^{-7}$ \\ 
PSTNN \cite{zhang2019infrared}(2019) & Patch size: 40, sliding step: 40, $\lambda=0.6/\sqrt{\mathrm{max}(n_1,n_2)\times n_3}$, $\epsilon=10^{-7}$                   \\
RLCM \cite{han2018infrared}(2018) & $k_1 = [2,5,9], k_2 = [4, 9, 16]$, scale: 3, threshold $k = 1$\\ 
ILCM \cite{han2020infrared}(2020) & Cell size: $3 \times 3$, threshold $k = 3$ \\
GSWLCM \cite{qiu2022global}(2022) & Local Window Structures: $[3,5,7,9]$, $\delta=0.01, k = 20$\\ 
\hline
\multicolumn{2}{l}{\textit{Data-Driven methods}}  \\ \hline 
ACM \cite{dai2021asymmetric}(2021)  &  Backbone: FPN, layer blocks: [4, 4, 4], channels: [8, 16, 32, 64]  \\ 
ALCNet \cite{dai2021attentional}(2021)        & Patch size: 40×40; Sliding step: 20; Attention threshold: 0.7; Loss weights: $\lambda_1$=0.6, $\lambda_2$=0.4         \\
DNANet \cite{li2022dense}(2022) & Backbone: resnet18, layer blocks: [2, 2, 2, 2], filter: [16, 32, 64, 128, 256]                   \\
ResUNet \cite{zhang2018road}(2018) & Initial Filters: 64; Kernel Size: 3×3 (all convolutional layers); Optimizer: Adam (lr = 0.001, $\beta_1$ = 0.9)     \\
RDIAN \cite{sun2023receptive}(2023) & RF Dilations: [1,2,4,8], Kernel=3×3; Optimizer: Adam (lr=1e-4, $\beta_1$=0.9, $\beta_2$=0.999)   \\
UIUNet \cite{wu2022uiu}(2023) & Channels: [64, 128, 256, 512], fuse mode: AsymBi                     \\
ISTDU-Net \cite{hou2021ristdnet}(2021)     &GSB Dilations: [1, 2, 3]; FAM Compression: r=16, Attention kernel=7×7;       \\
SCTransNet \cite{yuan2024sctransnet}(2024)  & Backbone: Swin-Tiny,  Layer Num=[2,6,6,2], Head Num=[3,6,12,24]\\
DATransNet \cite{hu2025datransnet}(2025)   & DeepSupervision: True, layer blocks: [2, 2, 2, 2], filter: [16, 32, 64, 128, 256]   \\ 

\rowcolor[rgb]{0.9,0.9,0.9}$\star$ \textbf{SWAN (Ours)} & DeepSupervision: True; Channels: [32, 64, 128, 256, 512], $M=16$ \\
\bottomrule[1pt]
\end{tabular}
\end{center}
\end{table*}


\section{Experiments and Analysis} \label{sec:experiment}

\subsection{Experimental Settings} \label{subsec:setting}

\subsubsection{Datasets} \label{subsec:datasets}
In our experiments, we utilized two datasets: IRSTD-Real and the NUDT dataset. We observed that the widely used public datasets IRSTD-1K and NUAA consist of real-world infrared imagery capturing the characteristic dimunitive nature of infrared small targets. Thus, we curated and merged these two datasets to form our novel IRSTD-Real dataset. The NUDT dataset serves as a synthetic dataset for comparative evaluation of experimental results.

\begin{figure}
    \centering
    \includegraphics[width=1\linewidth]{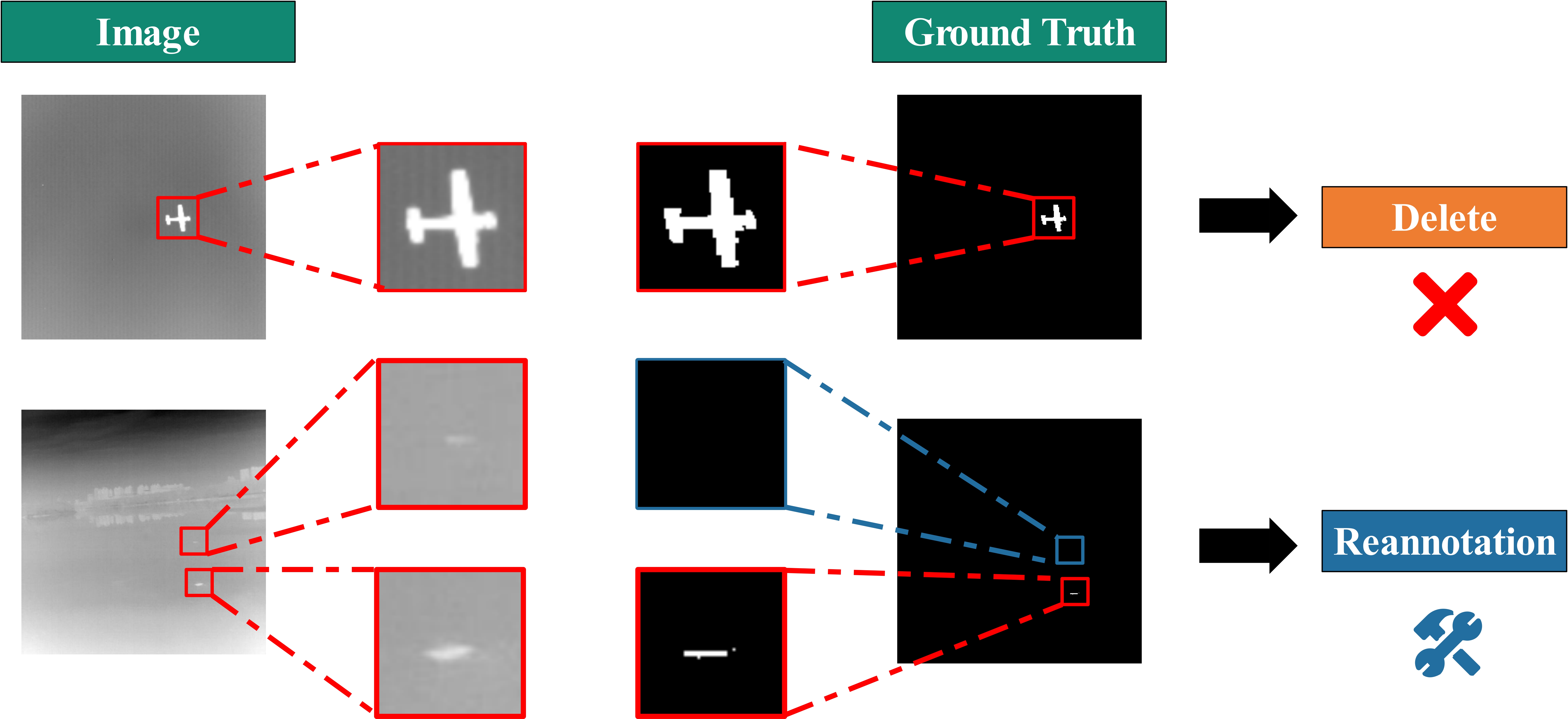}
    \caption{Images Excluded for Non-compliance with Small Target Criteria. Top: non-conforming small target entries pruned;
Bottom: label refinement and selective re-annotation performed.}
    \label{fig:datasets}
\end{figure}

\begin{enumerate}

\item{\textbf{IRSTD-Real}}: We refined the NUAA \cite{dai2021asymmetric} and IRSTD-1k \cite{zhang2023attention} datasets through optimization procedures: removing images inconsistent with small target criteria and re-annotating problematic labels (representative examples shown in the Fig. \ref{fig:datasets}). Subsequently, 1,200 rigorously curated images meeting small target standards comprise the final IRSTD-Real dataset, which contains 400 images at 256×256 resolution and 800 images at 512×512 resolution. This dataset encompasses targets (aircraft, vehicles) across diverse temporal conditions, weather patterns, and environments, featuring multi-scale small target instances—all annotated with pixel-level segmentation masks.


\item{\textbf{NUDT-SIRST}}: This dataset contains 1,327 synthetic images, all at a resolution of 256×256 pixels. It covers a variety of complex backgrounds, including skies, oceans, and urban environments, and is precisely annotated. This dataset can be generated on a large scale at a low cost and is widely used as a public resource for infrared small target detection.

\end{enumerate}

\subsubsection{Experimental Details}\label{subsec:details}


During the training phase, each dataset was partitioned into training and testing sets at an 8:2 ratio. To enhance the robustness and generalizability of our findings, this study incorporates two heterogeneous infrared datasets for cross-validation. Specifically, using UNet as the performance benchmark, we systematically conducted ablation studies on network modules using the IRSTD-Real dataset to rigorously evaluate the effectiveness and reliability of each component. Concurrently, Tab. \ref{tab:parameters}  provides a systematic comparison of core parameters (including method source and publication year) among current state-of-the-arts infrared small target detection (IRSTD) approaches that serve as comparable baselines. It is important to note that model-driven methods predominantly employ an unsupervised learning paradigm, while existing data-driven methods primarily fall within the supervised learning category.

All experiments were conducted over 400 epochs without using any pre-trained weights. Each image was normalized and randomly cropped to a size of $256 \times 256$ pixels. The following parameters: batch size of 16 and weight decay of 0.0005, the model's weights and biases were initialized using the Kaiming initialization method. The model was trained using the Binary Cross-Entropy (BCE) loss function and optimized with the Adam optimizer, starting with an initial learning rate of 0.001, which was gradually reduced to $1 \times 10^{-5}$ using a cosine annealing strategy. A fixed threshold of 0.5 was applied for segmenting significant maps. The experimental environment consisted of a vGPU with 32GB, a 16 vCPU AMD EPYC 9654 96-Core Processor, and the Ubuntu 18.04 operating system. The framework used was PyTorch 1.8.1, Python 3.8, and CUDA 11.1.

\begin{figure*}
    \centering
    \includegraphics[width=1\linewidth]{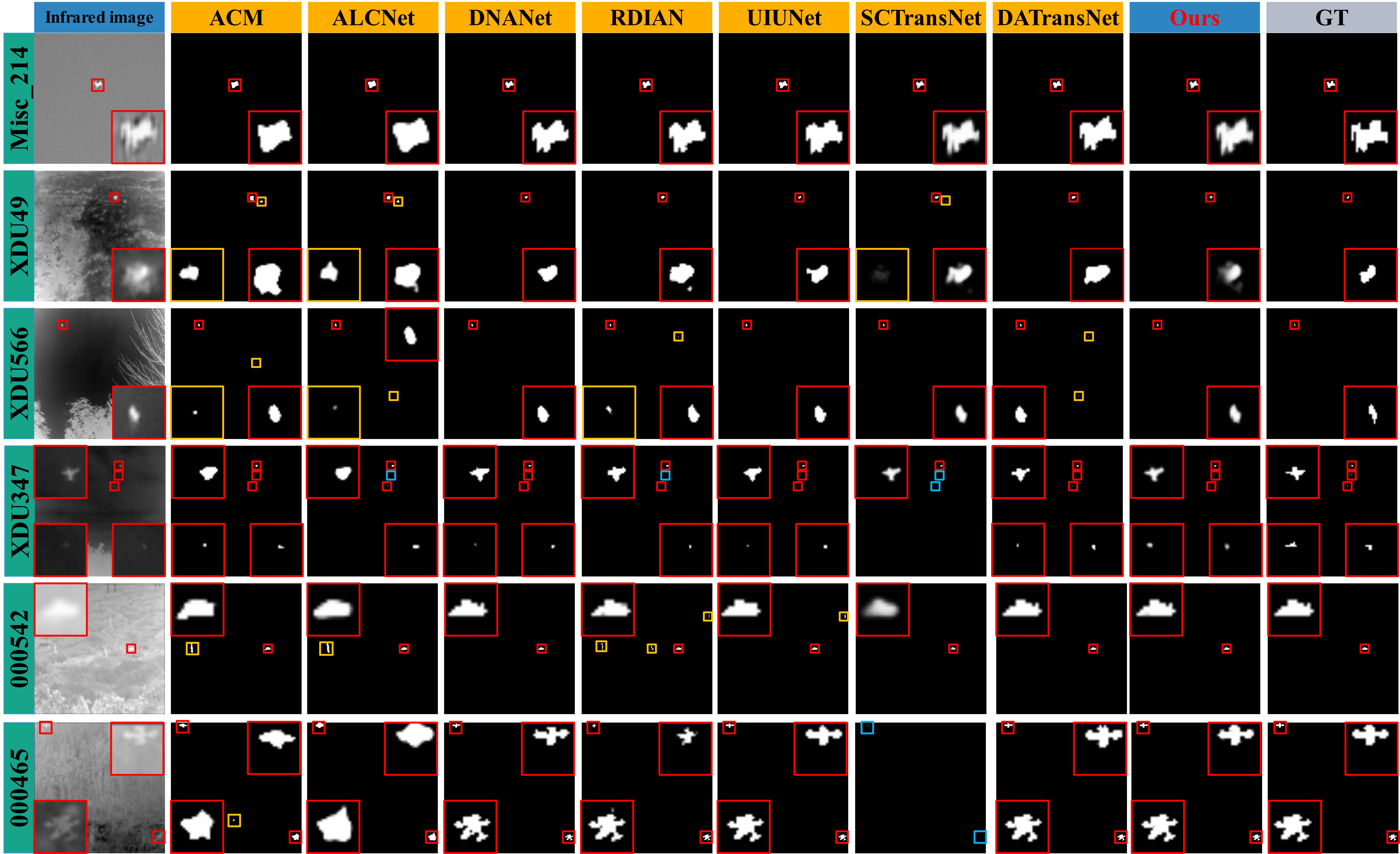}
    \caption{Visualizations of representative results are shown above. In the figures, false negatives (FNs) are marked with blue boxes, false positives (FPs) are indicated with orange boxes, and correctly detected targets (true positives / TPs) are highlighted with red boxes. Zoomed-in patches of the detection regions are displayed in the corners of each image.}
    \label{fig:Qualitative}
\end{figure*}

\subsubsection{Evaluation Metrics} \label{subsec:metrics}

To compare the proposed method with state-of-the-arts (SOTA) approaches, we employed commonly used evaluation metrics, including mIoU, nIoU, Pd, Fa, and F1. Additionally, we used parameters and flops to assess whether our model is more complex compared to the current SOTA models. Their definitions are as follows:

1) \textit{Mean Intersection over Union (mIoU)}:
\begin{equation}
\text{mIoU} = \frac{TP}{TP + FP + FN}
\label{eq:miou}
\end{equation}
Where TP, FP and TN, FN represent the number of correctly predicted positive samples, incorrectly predicted positive samples, correctly predicted negative samples, and incorrectly predicted negative samples.

2) \textit{Normalized Intersection over Union (nIoU)}: nIoU is the normalized version of IoU, given as
\begin{equation}
\text{nIoU} = \frac{1}{N} \sum_{i=1}^{N} \frac{TP(i)}{T(i) + P(i) - TP(i)}
\label{eq:niou}
\end{equation}
where N is the total number of samples, and T, P represent the number of positive pixels in the ground truth and predicted results, Unlike IoU, the improved metric for small target detection, nIoU, normalizes the calculations to reduce the influence of target size on IoU. This provides a better reflection of the algorithm's performance in low signal-to-noise ratio images.

3) \textit{Probability of Detection (${{P}_{d}}$)}: The probability of detecting actual targets, aimed at avoiding missed detections
\begin{equation}
  \text{Pd} = \frac{TP}{TP + FN}
  \label{eq:probability_of_detection}
\end{equation}

4) \textit{False-Alarm Rate (${{F}_{a}}$)}: \textit{${{F}_{a}}$} is the ratio of false predicted. FP  refers to the number of pixels that are incorrectly predicted as targets and ${FP + FN}$ represents the total number of incorrectly classified samples by the model.
\begin{equation}
    \text{Fa} = \frac{FP}{FP + TN} 
\end{equation}

5) \textit{F-measure (${{F}_{1}}$)}: This metric provides a balanced evaluation of the model’s performance in detecting targets while minimizing both missed detections and false alarms.
\begin{equation}
    F_1 = \frac{2TP}{2TP + FP + FN}.
\end{equation}
where ${TP}$, ${FP}$ and ${FN}$ denote the True Positive, False Positive and False Negative.

Additionally, the evaluation metrics for parameters and Flops are as follows: Parameters focus on the storage overhead and memory usage of the model, serving as indicators of the model's scale and capacity potential. Flops primarily assess the computational overhead of the model, representing the model's computational complexity and theoretical speed/energy consumption.

\subsection{Comparison with State-of-the-Arts} \label{subsec:sota}

\subsubsection{Quantitative comparison with State-of-the-Arts}

\begin{table*}[ht]
  \renewcommand\arraystretch{1.2}
  \footnotesize
  \centering
  \caption{Comparison with Other State-of-the-arts methods on two datasets. The $\uparrow$ indicates that the higher the indicator, the better. We display the best result in the \textcolor{red}{red} color and the second-best result in the \textcolor{blue}{blue} color, the third-best result in \textcolor{green}{green} color. Except for parameters and Flops, evaluation metrics use \% as the unit.}
  \label{tab:sota}
  \setlength{\tabcolsep}{3.0pt}
  \resizebox{\textwidth}{!}{%
  \begin{tabular}{c|c|c|ccccc|ccccc}
    \multirow{2}{*}{Method} & \multirow{2}{*}{Para$\downarrow$} & \multirow{2}{*}{Flops$\downarrow$}  & \multicolumn{5}{c|}{IRSTD-Real(Tr=80\%)}  & \multicolumn{5}{c}{NUDT(Tr=80\%)} \\
       & & &  mIoU $\uparrow$ & nIoU$\uparrow$ & Pd $\uparrow$ & Fa $\downarrow$ & F1 $\uparrow$ &  mIoU $\uparrow$ & nIoU$\uparrow$ & Pd $\uparrow$ & Fa $\downarrow$ & F1 $\uparrow$ \\
  \hline
   \multicolumn{13}{l}{\textit{Model-Driven methods}}  \\ 
  \hline
  FKRW \cite{qin2019infrared} &- &- & 20.14 & 26.18 & 83.25 & 19.57 & 18.17 & 13.25 & 19.89 & 76.37 & 65.34 & 23.17 \\ 
  IPI \cite{gao2013infrared} &- &- & 25.48 & 30.19 & 80.22 & 21.37 & 27.18 & 26.78 & 37.56 & 74.25 & 40.17 & 26.19  \\
  PSTNN \cite{zhang2019infrared} &- &- & 26.16 & 26.30 & 80.17  & 28.40 & 23.43  & 28.39  & 20.76 & 76.69 & 41.38  & 30.51  \\
  RLCM \cite{han2018infrared} &- &- & 25.84 & 30.75 & 71.78  & 30.18 & 36.92  & 14.88 & 5.98 & 72.00 & 60.38  & 38.36  \\
  ILCM \cite{han2020infrared} &- &- & 29.89 & 25.01 & 76.07  & 50.47 & 38.76 & 23.34 & 32.60  & 75.81 & 57.22 & 32.21  \\
  TLLCM \cite{han2019local} &- &- & 25.29 & 21.71 & 73.54  & 20.98 & 28.13 & 21.89 & 22.56  & 75.85 & 67.26 & 41.11  \\
  GSWLCM \cite{qiu2022global} &- &- & 27.94 & 29.36 & 72.19  & 23.49 & 40.17 & 29.76 & 25.67  & 72.47 & 54.18 & 39.89 \\
  \hline  
  \multicolumn{13}{l}{\textit{Data-Driven methods}}  \\ 
  \hline
  ACM \cite{dai2021asymmetric} & 0.40M &0.4G & 53.88 & 55.02 & 94.58  & 17.88 & 61.88  & 66.94 & 60.10  & 90.40 & 12.91 & 77.25  \\
  ALCNet \cite{dai2021attentional} &0.43M &0.38G & 71.54  & 57.76 & 93.06 & 21.67 &  64.82  & 62.70  & 68.54  & 92.25 & 10.13  & 65.64 \\
  DNANet \cite{li2022dense} & 4.7M & 14.26G & 70.22 & 61.54 & 94.33 & 17.65 & 68.69 & 91.28 & 78.79 & 93.26 & 5.39 & 75.58 \\
  ResUNet \cite{zhang2018road} &0.99M &3.83G & 65.81  & 59.69 & 95.11 & 17.05 &  76.83  & 84.42  & 85.78  & 95.18 & 7.80 & 72.80 \\
  RDIAN \cite{sun2023receptive} &0.22M &3.72G & 67.96  & 53.62 & 91.89 & 23.64 &  73.69  & 82.42  & 75.14  & 95.83 & 5.63 & 72.20 \\
  UIUNet \cite{wu2022uiu} &50.54M &54.43G & 65.49  & 59.57 & 95.09 & 18.38 & 76.71 & 91.62  & 87.81  & \textcolor{green}{96.74} & \textcolor{green}{3.10} & 78.69 \\
  ISTDU-Net \cite{hou2021ristdnet} &2.75M &7.94G & \textcolor{blue}{75.66}  & \textcolor{green}{69.64} & 95.23 & 17.71 & 79.77  & 93.08  & 88.21  & 95.52 & 3.63 & 80.05 \\
  SCTransNet \cite{yuan2024sctransnet} &11.19M &10.12G & 69.63  & 67.53 & \textcolor{green}{95.28}  & \textcolor{green}{16.70} & \textcolor{green}{81.86} & \textcolor{green}{94.09}  & \textcolor{green}{88.38}  & 93.62 & 4.29  & \textcolor{green}{86.95} \\
  DATransNet \cite{hu2025datransnet} & 2.18M & 8.18G & \textcolor{blue}{75.05} & \textcolor{blue}{70.54} &\textcolor{blue}{95.31}  & \textcolor{blue}{16.02} & \textcolor{red}{86.47} & \textcolor{red}{95.37} & \textcolor{blue}{90.67} & \textcolor{blue}{98.44} & \textcolor{blue}{2.26} & \textcolor{blue}{93.63} \\

  \hline
  \rowcolor[rgb]{0.9,0.9,0.9}$\star$ \textbf{SWAN(Ours)}  &5.48M &6.60G & \textcolor{red}{75.86}  & \textcolor{red}{72.42} & \textcolor{red}{95.67} & \textcolor{red}{14.48} & \textcolor{blue}{86.54}  & \textcolor{blue}{94.13}  & \textcolor{red}{93.79}  & \textcolor{red}{98.84} & \textcolor{red}{2.13} & \textcolor{red}{94.03}  \\
  \end{tabular}
  }
\end{table*}

In this section, we compared the sota IRSTD methods (ACM, ALCNet, DNANet, RDIAN, UIUNet, SCTransNet, DATransNet) on two datasets.

Tab. \ref{tab:sota} presents the quantitative comparison results compared with the sota method. The method demonstrated high performance on all evaluation metrics on the IRSTD-Real and NUDT datasets, proving the effectiveness of this method. On the IRSTD-Real dataset, our method SWAN achieved good results, with Pd (\%) of 95.67, Fa ($10^{-6}$) of 14.48, mIoU (\%) of 75.86 and nIoU (\%) of 72.42. On the NUDT dataset, Pd (\%) was 98.84 and Fa ($10^{-6}$) of 2.13, mIoU (\%) 94.13 and nIoU (\%) 93.79 achieved the best results compared with ACM, DNANet, ALCNet, UIUNet and SCTransNet, DATransNet. It is worth noting that our SWAN has achieved the best mIoU metric on the IRSTD-Real dataset. This performance can be attributed to the division and interaction of high and low frequencies, as well as the effectiveness of the frequency-domain spatial fusion strategy, which eliminates false alarms while retaining weak and small targets and helps identify the influence of small targets and noise. The comparative experiments show that the data-driven method based on deep learning maintains high detection accuracy due to its ability to extract semantic information and the minimum dependence on hyperparameters, thereby enhancing the robustness to scene changes.

\begin{figure*}
    \centering
    \includegraphics[width=1\linewidth]{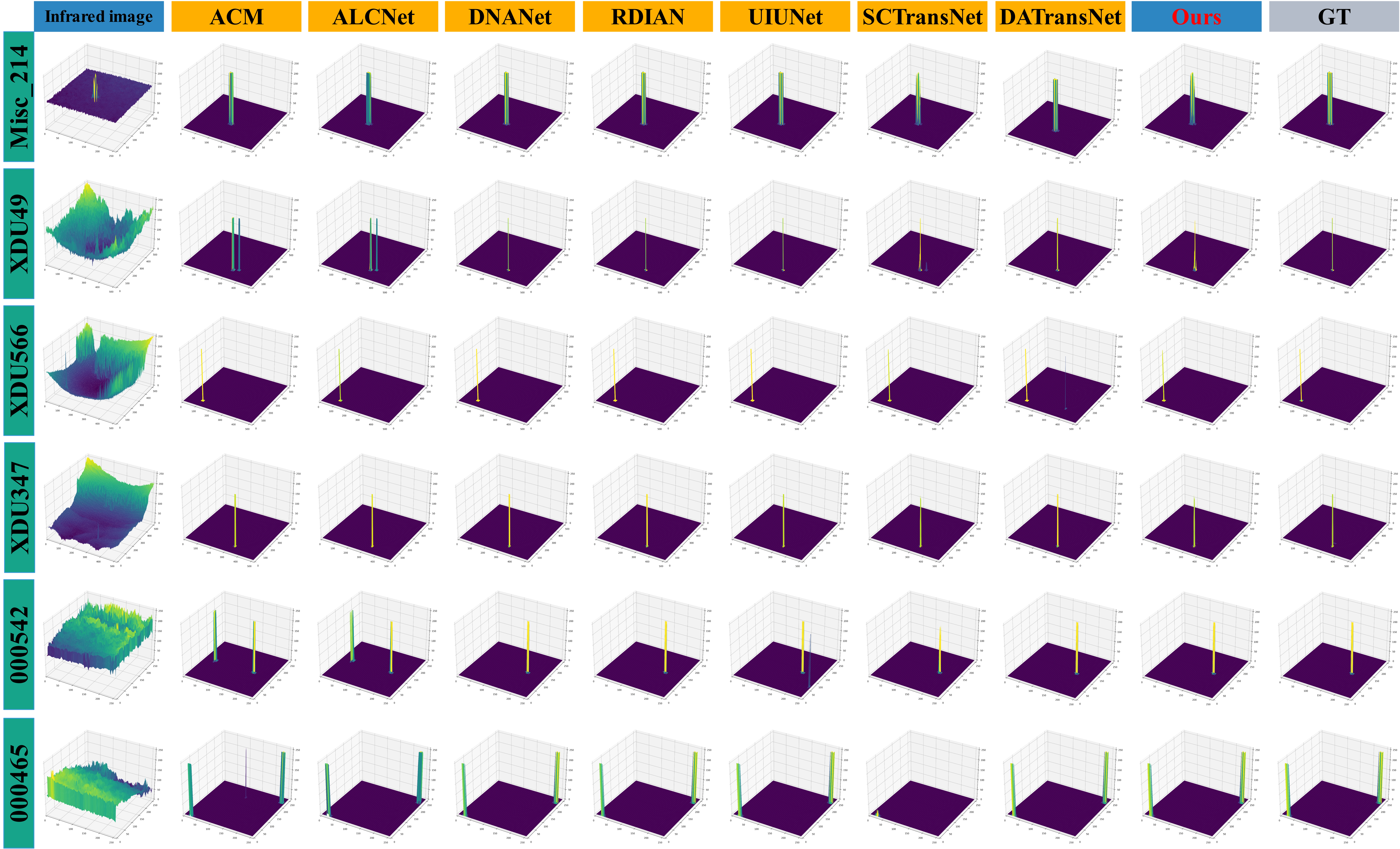}
    \caption{3D visualization results of different methods on 6 test images.}
    \label{fig:3D}
\end{figure*}

Our method is always superior to the SCTransNet. Although it integrates local edge information in the spatial domain, it lacks accuracy in frequency-domain feature extraction, high-frequency and low-frequency information exchange, and dynamic fusion in the frequency-domain space. These features ensure the fine segmentation of infrared small targets and achieve good results. The 3D visualization parameters of the IRSTD-Real and NUDT datasets (as shown in Fig. \ref{fig:3D}) confirm the results in Tab. \ref{tab:sota} and the stability and superiority of the deep learning methods.

\begin{figure*}[]
  \centering
  \subfloat[IRSTD-Real]{\includegraphics[width=0.48\textwidth]{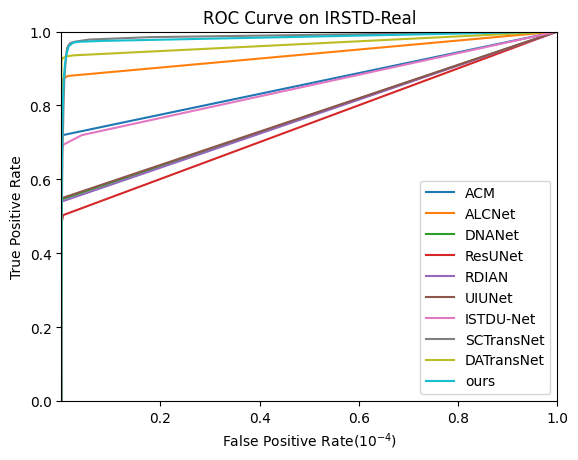}}
  \hfill
  \subfloat[NUDT]{\includegraphics[width=0.48\textwidth]{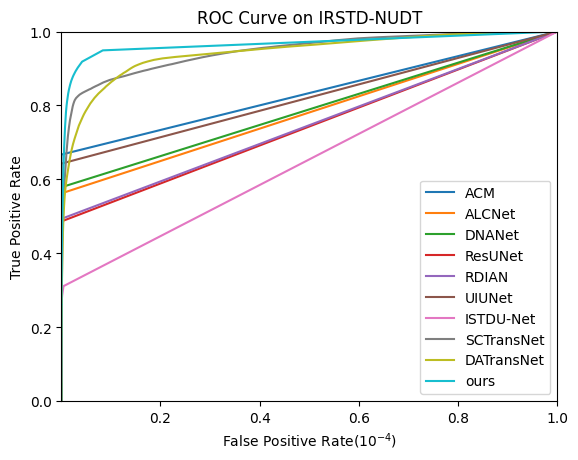}}
  \caption{ROC curves of different methods on the IRSTD-Real and NUDT datasets. Our SWAN achieves the highest Pd at very low Fa.}
  \label{fig:roc}
\end{figure*}

Fig. \ref{fig:roc} shows the ROC curves comparing our method with other SOTA methods on the IRSTD-Real dataset. SWAN achieved the highest mIoU and nIoU, demonstrating its strong detection capability. Notably, our network maintains a favorable balance between Pd and Fa, ensuring stable performance across various scenes. This reflects its limited robustness in complex infrared scenarios.

\subsubsection{Qualitative Visualization and Analysis} \label{subsec:visualization}

We selected pictures from the two datasets for visual comparison by different methods, as shown in Fig. \ref{fig:Qualitative}. The correctly detected targets, Fa and the areas corresponding to missed detections are displayed in red, orange and blue boxes. Compared with the large-scale false detection and missed detection behaviors of traditional methods, the object segmentation based on deep learning can distinguish the object from the background more accurately. Currently, more difficulties are concentrated on how to precisely segment the object edge. As shown in the sixth row 000465 of the figure, our method can segment the object edge more accurately. Other methods all have segmentation errors to varying degrees. The missed detection of SCTransNet, the edge segmentation errors of ACM and ALCNet may all lead to losses of varying degrees in practical applications.


\subsection{Ablation Study} \label{subsec:ablation}
\subsubsection{\textbf{Ablation of Different Model-wise Components}} \label{subsec:lamablation}

\begin{table*}[ht]
  \setlength{\abovecaptionskip}{0cm}  
  \renewcommand\arraystretch{1.2}
  \footnotesize
  \centering
  \vspace{-1\baselineskip}
  \caption{Quantitative Ablation Study on SWAN( Haar Wavelet Convolution \textbf{HWConv},  Shifted Window Self-Attentio is mared \textbf{SSA}, Residual Dual-Channel Attention is mared \textbf{RDCA})}
  \label{tab: SWAN module ablation}
  \setlength{\tabcolsep}{3.0pt}
  \resizebox{\textwidth}{!}{%
  \begin{tabular}{c|ccc|ccccc|ccccc}
      \multirow{2}{*}{Ablation} & \multicolumn{3}{c|}{Module} & \multicolumn{5}{c|}{IRSTD-Real} & \multicolumn{5}{c}{NUDT} \\
      & \textbf{HWConv} & \textbf{SSA} & \textbf{RDCA}   & mIoU $\uparrow$ &  nIoU  $\uparrow$ & Pd $\uparrow$ &  Fa $\downarrow$ &  F1 $\uparrow$ & mIoU $\uparrow$ &  nIoU  $\uparrow$ & Pd $\uparrow$ &  Fa $\downarrow$ &  F1$\uparrow$ \\
  \Xhline{1pt}
  (a) & \xmark & \xmark & \xmark & 62.88 & 68.60 & 89.10 &  23.31 &  75.70 &  81.53 & 75.81 & 94.92 & 9.65 & 92.83 \\
  (b) & $\checkmark$ & \xmark & \xmark & 68.22 & 70.55 & 92.62 &  20.33 &  77.71 & 84.60 & 85.24 & 95.18 & 4.77 & 92.29 \\
  (c) & $\checkmark$ & $\checkmark$ & \xmark & 72.25 & 71.62 & 94.25 & 18.52 & 84.67 & 87.34  & 89.12 &97.72 & 5.79 & 93.21  \\ 
  \hline
  \rowcolor[rgb]{0.9,0.9,0.9}
  (d) & \textbf{$\checkmark$} & $\checkmark$ & $\checkmark$ & \textbf{75.86}  & \textbf{72.42} & \textbf{95.67} & \textbf{14.48} & \textbf{86.54} & \textbf{94.13} & \textbf{93.79} & \textbf{98.84} & \textbf{2.13} & \textbf{94.03} \\
  
\end{tabular}
}
\end{table*}


We chose UNet as our baseline to validate the different impacts of the HWConv, SSA, and RDCA modules. The experimental results indicate that all three modules can serve as plug-and-play components to enhance the evaluation results of the network model.

The specific results are shown in Tab. \ref{tab: SWAN module ablation}
\begin{enumerate}

\item Each module improves segmentation performance relative to the baseline. The standards for enhancing cross-domain integration and segmentation performance are met.

\item As each module is added, it can be observed that both nIoU and Pd have risen compared to the baseline model, while Fa has decreased.

\item On the NUDT synthetic dataset, the increase in Fa after adding SSA may be due to the spatial information interaction primarily affecting IRSTD-Real datasets, as the synthetic dataset has disrupted the original spatial environment.

\end{enumerate}

\begin{figure}
    \centering
    \includegraphics[width=1\linewidth]{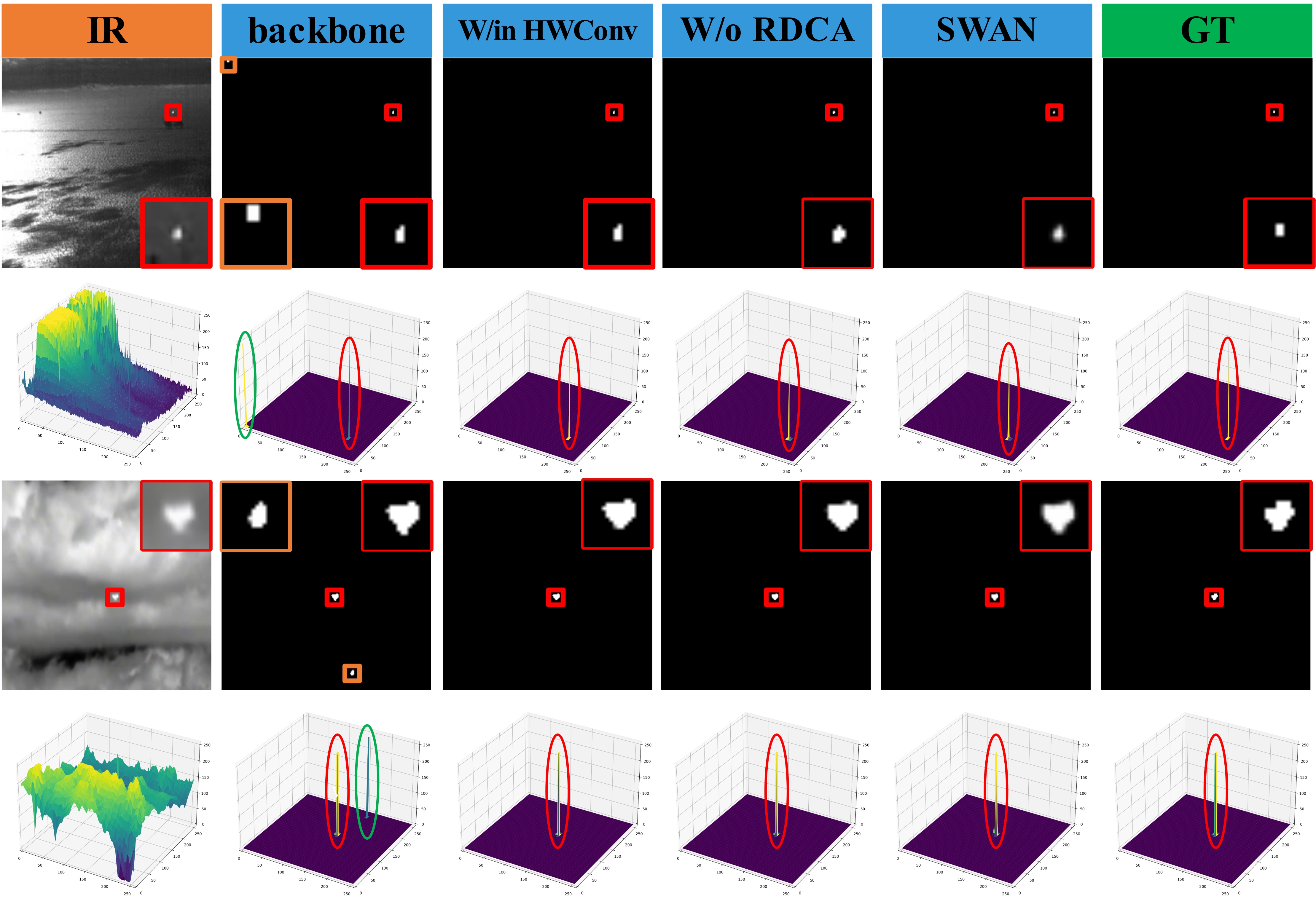}
    \caption{Visualization of ablation study results: (a) Detection results of backbone. (b) Detection results of backbone \& HWConv module. (c) Detection results without RDCA module. (d) Detection results of SWAN.}
    \label{fig:ablation}
\end{figure}

As illustrated in the Fig. \ref{fig:ablation}, the visualization heatmaps further validate the effectiveness of individual modules and their combinations.

\subsubsection{\textbf{Ablation of HWConv wavelets level}} \label{subsec:lamablation-1}

\begin{table}[ht]
  \setlength{\abovecaptionskip}{0cm} 
  \renewcommand\arraystretch{1.2}
  \footnotesize
  \centering
  \vspace{-1\baselineskip}
  \caption{Ablation Study on levels of nested layers.}
  \label{tab:levels}
  \setlength{\tabcolsep}{1pt}
  \resizebox{\columnwidth}{!}{ 
    \begin{tabular}{c|ccccc}
      \multirow{2}{*}{nest level} & \multicolumn{5}{c}{Evaluation Metrics}\\
      & mIoU $\uparrow$ & nIoU  $\uparrow$  & Pd $\uparrow$ & Fa $\downarrow$ & F1 $\uparrow$\\
      \Xhline{1pt}
      1 & 75.72 & 72.16 & 94.92 & 14.29 & 86.15 \\
      \hline
      \rowcolor[rgb]{0.9,0.9,0.9}
      \textbf{2} & \textbf{75.86} & \textbf{72.42} & \textbf{95.67} & \textbf{14.48} & \textbf{86.54} \\
      \hline
      3 & 75.52 & 71.62 & 95.41 & 14.21 & 86.36 \\
      4 & 75.47 & 70.28 & 95.20 & 14.43 & 85.31 \\
      5 & 74.86 & 69.42 &92.44 & 20.15 & 84.36 \\
    \end{tabular}
}
\end{table}


Through systematic ablation experiments of five nested levels in the HWConv architecture conducted on the IRSTD-Real dataset, it was found that the model achieves optimal performance on comprehensive evaluation metrics when the nesting level is set to two. As shown in Tab. \ref{tab:levels}, key evaluation metrics including mIoU, nIoU, Pd, Fa and the F1 score exhibited significant improvements. Although a slight increase in the Fa was observed, the magnitude of this increase remained within acceptable limits and did not substantially compromise the model's overall effectiveness. The experiments further revealed that as the nesting level increased beyond two (>2 levels), the model's evaluation metrics generally exhibited a declining trend.

These results indicate that the two-level nested structure achieves an optimal balance between feature abstraction capability and computational efficiency: the first level (L1) focuses on efficient extraction of local features, while the second level (L2) effectively facilitates cross-modal interaction and fusion. Notably, when the nesting depth exceeds two levels (>L2), model performance not only tends to saturate but even demonstrates degradation. This phenomenon is primarily attributed to the additional computational burden introduced by layer redundancy, which fails to deliver commensurate performance gains, resulting in diminishing marginal returns.

\subsubsection{\textbf{Ablation of Different wavelets in HWConv}} \label{subsec:lamablation-2}

\begin{table}[]
  \setlength{\abovecaptionskip}{0cm} 
  \renewcommand\arraystretch{1.2}
  \footnotesize
  \centering
  \vspace{-1\baselineskip}
  \caption{Ablation Study on different wavelets.}
  \label{tab:wavelets}
  \setlength{\tabcolsep}{1pt}
  \resizebox{\columnwidth}{!}{ 
  \begin{tabular}{c|ccccc}
  \multirow{2}{*}{wavelets} & \multicolumn{5}{c}{Evaluation Metrics}\\
       & mIoU $\uparrow$ &  nIoU  $\uparrow$ &  Pd $\uparrow$ &  Fa $\downarrow$  & F1 $\uparrow$ \\
  \Xhline{1pt}
  Symlet & 75.14 & 72.18 & 94.19 & 17.51 & 85.02 \\
  Coiflet & 75.38 & 71.83 & 94.38 & 15.95 & 87.27  \\
  Biorthogonal & 75.19 & 70.49 & 94.67 & 16.62 & 86.15  \\
  Reverse Biorthogonal & 75.45 & 71.39 & 95.68 & 14.50 & 86.29 \\
  \hline
  \rowcolor[rgb]{0.9,0.9,0.9}
  \textbf{Haar} & \textbf{75.86}  & \textbf{72.42} & \textbf{95.67} & \textbf{14.48} & \textbf{86.54} \\
  
\end{tabular}
 }
\end{table}
To optimize the feature extraction capability of HWConv, We conduct a systematic ablation study in the IRSTD-Real dataset with six wavelet basis functions (Haar\cite{rosenfeld1976digital}, Symlet\cite{unser1995texture}, Coiflet\cite{do2002wavelet}, Biorthogonal\cite{cohen1992biorthogonal}, Reverse Biorth \-ogonal\cite{shensa2002discrete}). The comprehensive evaluation metrics (mIoU, nIoU, Pd, Fa, F1) indicate that db1 (Haar wavelet) significantly outperforms the other wavelet bases (as shown in Tab. \ref{tab:wavelets}). Specifically, db1 leads in key metrics: mIoU: 75.86\%, F1: 86.54\%, Noise suppression capability is outstanding, with a low false alarm rate (Fa) of 14.48\%, Edge feature retention is advantageous, achieving an nIoU (narrowband Intersection over Union) of 72.42\%. The experiments also reveal that while complex wavelet bases (such as Coiflet) show slight gains in local texture representation (F1↑ 0.73\%), their introduction of frequency band aliasing results in a lower Pd (↓1.29\%), ultimately reducing the model's robustness. Thanks to its tight support and computational efficiency, Haar emerges as the optimal choice.


\section{Conclusion} \label{sec:conclusion}


A new infrared small target detection method has been proposed Synergistic Wavelet-Attention Network (SWAN). This method effectively utilizes frequency domain characteristics and convolution information, introducing long-range feature dependencies at different scales. It enhances the target self-attention expression while adaptively processing both deep and shallow feature information to obtain the final fusion result. The HWConv employs Haar wavelet transformation to retain the smooth information and overall structure of the original image, combined with convolution to preserve texture features. The SSA module interacts with the spatial positional information of the extracted features, thereby improving the performance of feature information transmission in space. Finally, the RDCA adjusts the weights of features at different levels, enhancing the overall target segmentation performance. We evaluated existing state-of-the-arts methods. Experimental results demonstrate that SWAN achieved favorable outcomes, validating the superiority of this method. Additionally, an ablation study of the modules was conducted, fully verifying the robustness and noise resistance of the algorithm.

\bibliographystyle{IEEEtran}
\bibliography{./reference.bib}

\begin{thebibliography}{10}
\providecommand{\url}[1]{#1}
\csname url@samestyle\endcsname
\providecommand{\newblock}{\relax}
\providecommand{\bibinfo}[2]{#2}
\providecommand{\BIBentrySTDinterwordspacing}{\spaceskip=0pt\relax}
\providecommand{\BIBentryALTinterwordstretchfactor}{4}
\providecommand{\BIBentryALTinterwordspacing}{\spaceskip=\fontdimen2\font plus
\BIBentryALTinterwordstretchfactor\fontdimen3\font minus \fontdimen4\font\relax}
\providecommand{\BIBforeignlanguage}[2]{{%
\expandafter\ifx\csname l@#1\endcsname\relax
\typeout{** WARNING: IEEEtran.bst: No hyphenation pattern has been}%
\typeout{** loaded for the language `#1'. Using the pattern for}%
\typeout{** the default language instead.}%
\else
\language=\csname l@#1\endcsname
\fi
#2}}
\providecommand{\BIBdecl}{\relax}
\BIBdecl

\bibitem{sobrino2016review}
J.~A. Sobrino, F.~Del~Frate, M.~Drusch, J.~C. Jim{\'e}nez-Mu{\~n}oz, P.~Manunta, and A.~Regan, ``Review of thermal infrared applications and requirements for future high-resolution sensors,'' \emph{IEEE Transactions on Geoscience and Remote Sensing}, vol.~54, no.~5, pp. 2963--2972, 2016.

\bibitem{deng2016small}
H.~Deng, X.~Sun, M.~Liu, C.~Ye, and X.~Zhou, ``Small infrared target detection based on weighted local difference measure,'' \emph{IEEE Transactions on Geoscience and Remote Sensing}, vol.~54, no.~7, pp. 4204--4214, 2016.

\bibitem{wu2023mtu}
T.~Wu, B.~Li, Y.~Luo, Y.~Wang, C.~Xiao, T.~Liu, J.~Yang, W.~An, and Y.~Guo, ``Mtu-net: Multilevel transunet for space-based infrared tiny ship detection,'' \emph{IEEE Transactions on Geoscience and Remote Sensing}, vol.~61, pp. 1--15, 2023.

\bibitem{hu2024smpisd}
C.~Hu, X.~Dong, Y.~Huang, L.~Wang, L.~Xu, T.~Pu, and Z.~Peng, ``Smpisd-mtpnet: Scene semantic prior-assisted infrared ship detection using multi-task perception networks,'' \emph{IEEE Transactions on Geoscience and Remote Sensing}, 2024.

\bibitem{sun2020infrared}
Y.~Sun, J.~Yang, and W.~An, ``Infrared dim and small target detection via multiple subspace learning and spatial-temporal patch-tensor model,'' \emph{IEEE Transactions on Geoscience and Remote Sensing}, vol.~59, no.~5, pp. 3737--3752, 2020.

\bibitem{zhu2024towards}
Y.~Zhu, Y.~Ma, F.~Fan, J.~Huang, K.~Wu, and G.~Wang, ``Towards accurate infrared small target detection via edge-aware gated transformer,'' \emph{IEEE Journal of Selected Topics in Applied Earth Observations and Remote Sensing}, 2024.

\bibitem{zhu2020balanced}
H.~Zhu, J.~Zhang, G.~Xu, and L.~Deng, ``Balanced ring top-hat transformation for infrared small-target detection with guided filter kernel,'' \emph{IEEE Transactions on Aerospace and Electronic Systems}, vol.~56, no.~5, pp. 3892--3903, 2020.

\bibitem{lu2022enhanced}
Z.~Lu, Z.~Huang, Q.~Song, K.~Bai, and Z.~Li, ``An enhanced image patch tensor decomposition for infrared small target detection,'' \emph{Remote Sensing}, vol.~14, no.~23, p. 6044, 2022.

\bibitem{deng2021entropy}
L.~Deng, G.~Xu, J.~Zhang, and H.~Zhu, ``Entropy-driven morphological top-hat transformation for infrared small target detection,'' \emph{IEEE Transactions on Aerospace and Electronic Systems}, vol.~58, no.~2, pp. 962--975, 2021.

\bibitem{deng2021infrared}
L.~Deng, J.~Zhang, G.~Xu, and H.~Zhu, ``Infrared small target detection via adaptive m-estimator ring top-hat transformation,'' \emph{Pattern Recognition}, vol. 112, p. 107729, 2021.

\bibitem{bai2010analysis}
X.~Bai and F.~Zhou, ``Analysis of new top-hat transformation and the application for infrared dim small target detection,'' \emph{Pattern Recognition}, vol.~43, no.~6, pp. 2145--2156, 2010.

\bibitem{li2021infrared}
H.~Li, Q.~Wang, H.~Wang, and W.~Yang, ``Infrared small target detection using tensor based least mean square,'' \emph{Computers \& Electrical Engineering}, vol.~91, p. 106994, 2021.

\bibitem{zhang2023infrared}
X.~Zhang, J.~Ru, and C.~Wu, ``Infrared small target detection based on gradient correlation filtering and contrast measurement,'' \emph{IEEE Transactions on Geoscience and Remote Sensing}, vol.~61, pp. 1--12, 2023.

\bibitem{ren2020infrared}
K.~Ren, C.~Song, X.~Miao, M.~Wan, J.~Xiao, G.~Gu, and Q.~Chen, ``Infrared small target detection based on non-subsampled shearlet transform and phase spectrum of quaternion fourier transform,'' \emph{Optical and Quantum Electronics}, vol.~52, pp. 1--15, 2020.

\bibitem{chen2013local}
C.~P. Chen, H.~Li, Y.~Wei, T.~Xia, and Y.~Y. Tang, ``A local contrast method for small infrared target detection,'' \emph{IEEE transactions on geoscience and remote sensing}, vol.~52, no.~1, pp. 574--581, 2013.

\bibitem{wei2016multiscale}
Y.~Wei, X.~You, and H.~Li, ``Multiscale patch-based contrast measure for small infrared target detection,'' \emph{Pattern Recognition}, vol.~58, pp. 216--226, 2016.

\bibitem{qiu2022global}
Z.~Qiu, Y.~Ma, F.~Fan, J.~Huang, and L.~Wu, ``Global sparsity-weighted local contrast measure for infrared small target detection,'' \emph{IEEE Geoscience and Remote Sensing Letters}, vol.~19, pp. 1--5, 2022.

\bibitem{qiu2020adaptive}
Z.~Qiu, Y.~Ma, F.~Fan, J.~Huang, and M.~Wu, ``Adaptive scale patch-based contrast measure for dim and small infrared target detection,'' \emph{IEEE Geoscience and Remote Sensing Letters}, vol.~19, pp. 1--5, 2020.

\bibitem{xu2023infrared}
Y.~Xu, M.~Wan, X.~Zhang, J.~Wu, Y.~Chen, Q.~Chen, and G.~Gu, ``Infrared small target detection based on local contrast-weighted multidirectional derivative,'' \emph{IEEE Transactions on Geoscience and Remote Sensing}, vol.~61, pp. 1--16, 2023.

\bibitem{lu2023infrared}
Z.~Lu, Z.~Huang, Q.~Song, H.~Ni, and K.~Bai, ``Infrared small target detection based on joint local contrast measures,'' \emph{Optik}, vol. 273, p. 170437, 2023.

\bibitem{dai2017reweighted}
Y.~Dai and Y.~Wu, ``Reweighted infrared patch-tensor model with both nonlocal and local priors for single-frame small target detection,'' \emph{IEEE journal of selected topics in applied earth observations and remote sensing}, vol.~10, no.~8, pp. 3752--3767, 2017.

\bibitem{zhang2018infrared}
L.~Zhang, L.~Peng, T.~Zhang, S.~Cao, and Z.~Peng, ``Infrared small target detection via non-convex rank approximation minimization joint l 2, 1 norm,'' \emph{Remote Sensing}, vol.~10, no.~11, p. 1821, 2018.

\bibitem{zhang2019infrared}
L.~Zhang and Z.~Peng, ``Infrared small target detection based on partial sum of the tensor nuclear norm,'' \emph{Remote Sensing}, vol.~11, no.~4, p. 382, 2019.

\bibitem{zhong2023infrared}
S.~Zhong, H.~Zhou, X.~Cui, X.~Cao, F.~Zhang \emph{et~al.}, ``Infrared small target detection based on local-image construction and maximum correntropy,'' \emph{Measurement}, vol. 211, p. 112662, 2023.

\bibitem{li2022dense}
B.~Li, C.~Xiao, L.~Wang, Y.~Wang, Z.~Lin, M.~Li, W.~An, and Y.~Guo, ``Dense nested attention network for infrared small target detection,'' \emph{IEEE Transactions on Image Processing}, vol.~32, pp. 1745--1758, 2022.

\bibitem{zhang2022isnet}
M.~Zhang, R.~Zhang, Y.~Yang, H.~Bai, J.~Zhang, and J.~Guo, ``{ISNET}: Shape matters for infrared small target detection,'' in \emph{Proceedings of the IEEE/CVF Conference on Computer Vision and Pattern Recognition}, 2022, pp. 877--886.

\bibitem{dai2021attentional}
Y.~Dai, Y.~Wu, F.~Zhou, and K.~Barnard, ``Attentional local contrast networks for infrared small target detection,'' \emph{IEEE transactions on geoscience and remote sensing}, vol.~59, no.~11, pp. 9813--9824, 2021.

\bibitem{zhang2023attention}
T.~Zhang, L.~Li, S.~Cao, T.~Pu, and Z.~Peng, ``Attention-guided pyramid context networks for detecting infrared small target under complex background,'' \emph{IEEE Transactions on Aerospace and Electronic Systems}, vol.~59, no.~4, pp. 4250--4261, 2023.

\bibitem{bai2022cross}
Y.~Bai, R.~Li, S.~Gou, C.~Zhang, Y.~Chen, and Z.~Zheng, ``Cross-connected bidirectional pyramid network for infrared small-dim target detection,'' \emph{IEEE Geoscience and Remote Sensing Letters}, vol.~19, pp. 1--5, 2022.

\bibitem{wu2022uiu}
X.~Wu, D.~Hong, and J.~Chanussot, ``Uiu-net: U-net in u-net for infrared small object detection,'' \emph{IEEE Transactions on Image Processing}, vol.~32, pp. 364--376, 2022.

\bibitem{ronneberger2015u}
O.~Ronneberger, P.~Fischer, and T.~Brox, ``U-net: Convolutional networks for biomedical image segmentation,'' in \emph{Medical image computing and computer-assisted intervention--MICCAI 2015: 18th international conference, Munich, Germany, October 5-9, 2015, proceedings, part III 18}.\hskip 1em plus 0.5em minus 0.4em\relax Springer, 2015, pp. 234--241.

\bibitem{hu2025datransnet}
C.~Hu, Y.~Huang, K.~Li, L.~Zhang, C.~Long, Y.~Zhu, T.~Pu, and Z.~Peng, ``Datransnet: dynamic attention transformer network for infrared small target detection,'' \emph{IEEE Geoscience and Remote Sensing Letters}, 2025.

\bibitem{zhong2022detecting}
Y.~Zhong, B.~Li, L.~Tang, S.~Kuang, S.~Wu, and S.~Ding, ``Detecting camouflaged object in frequency domain,'' in \emph{Proceedings of the IEEE/CVF Conference on Computer Vision and Pattern Recognition (CVPR)}, June 2022, pp. 4504--4513.

\bibitem{liu2023infrared}
F.~Liu, C.~Gao, F.~Chen, D.~Meng, W.~Zuo, and X.~Gao, ``Infrared small and dim target detection with transformer under complex backgrounds,'' \emph{IEEE Transactions on Image Processing}, vol.~32, pp. 5921--5932, 2023.

\bibitem{zhang2025exploring}
T.~Zhang, Y.~Zhu, J.~Zhao, G.~Cui, and Y.~Zheng, ``Exploring state space model in wavelet domain: An infrared and visible image fusion network via wavelet transform and state space model,'' \emph{arXiv preprint arXiv:2503.18378}, 2025.

\bibitem{yuan2024sctransnet}
S.~Yuan, H.~Qin, X.~Yan, N.~Akhtar, and A.~Mian, ``Sctransnet: Spatial-channel cross transformer network for infrared small target detection,'' \emph{IEEE Transactions on Geoscience and Remote Sensing}, 2024.

\bibitem{han2018infrared}
J.~Han, K.~Liang, B.~Zhou, X.~Zhu, J.~Zhao, and L.~Zhao, ``Infrared small target detection utilizing the multiscale relative local contrast measure,'' \emph{IEEE Geoscience and Remote Sensing Letters}, vol.~15, no.~4, pp. 612--616, 2018.

\bibitem{gao2013infrared}
C.~Gao, D.~Meng, Y.~Yang, Y.~Wang, X.~Zhou, and A.~G. Hauptmann, ``Infrared patch-image model for small target detection in a single image,'' \emph{IEEE transactions on image processing}, vol.~22, no.~12, pp. 4996--5009, 2013.

\bibitem{dai2021asymmetric}
Y.~Dai, Y.~Wu, F.~Zhou, and K.~Barnard, ``Asymmetric contextual modulation for infrared small target detection,'' in \emph{Proceedings of the IEEE/CVF winter conference on applications of computer vision}, 2021, pp. 950--959.

\bibitem{xiao2024background}
M.~Xiao, Q.~Dai, Y.~Zhu, K.~Guo, H.~Wang, X.~Shu, J.~Yang, and Y.~Dai, ``Background semantics matter: Cross-task feature exchange network for clustered infrared small target detection with sky-annotated dataset,'' \emph{arXiv preprint arXiv:2407.20078}, 2024.

\bibitem{zhu2021dau}
Y.~Zhu, S.~Tang, Y.~Jiang, and R.~Kang, ``Dau-net: A regression cell counting method,'' in \emph{ISCTT 2021; 6th International Conference on Information Science, Computer Technology and Transportation}.\hskip 1em plus 0.5em minus 0.4em\relax VDE, 2021, pp. 1--6.

\bibitem{zhang2025wife}
T.~Zhang, J.~Zhao, Y.~Zhu, and G.~Cui, ``Wife-fusion: Wavelet-aware intra-inter frequency enhancement for multi-model image fusion,'' \emph{arXiv preprint arXiv:2506.03555}, 2025.

\bibitem{yang2023wavecnns}
W.~Yang, B.~Chen, Y.~Shen, and L.~Yu, ``Wavecnns-at: Wavelet-based deep cnns of adaptive threshold for signal recognition,'' \emph{Applied Intelligence}, vol.~53, no.~23, pp. 28\,819--28\,831, 2023.

\bibitem{zhao2022wavelet}
X.~Zhao, P.~Huang, and X.~Shu, ``Wavelet-attention cnn for image classification,'' \emph{Multimedia Systems}, vol.~28, no.~3, pp. 915--924, 2022.

\bibitem{hu2018squeeze}
J.~Hu, L.~Shen, and G.~Sun, ``Squeeze-and-excitation networks,'' in \emph{Proceedings of the IEEE conference on computer vision and pattern recognition}, 2018, pp. 7132--7141.

\bibitem{woo2018cbam}
S.~Woo, J.~Park, J.-Y. Lee, and I.~S. Kweon, ``Cbam: Convolutional block attention module,'' in \emph{Proceedings of the European conference on computer vision (ECCV)}, 2018, pp. 3--19.

\bibitem{vaswani2017attention}
A.~Vaswani, N.~Shazeer, N.~Parmar, J.~Uszkoreit, L.~Jones, A.~N. Gomez, {\L}.~Kaiser, and I.~Polosukhin, ``Attention is all you need,'' \emph{Advances in neural information processing systems}, vol.~30, 2017.

\bibitem{liu2021swin}
Z.~Liu, Y.~Lin, Y.~Cao, H.~Hu, Y.~Wei, Z.~Zhang, S.~Lin, and B.~Guo, ``Swin transformer: Hierarchical vision transformer using shifted windows,'' in \emph{Proceedings of the IEEE/CVF international conference on computer vision}, 2021, pp. 10\,012--10\,022.

\bibitem{finder2024wavelet}
S.~E. Finder, R.~Amoyal, E.~Treister, and O.~Freifeld, ``Wavelet convolutions for large receptive fields,'' in \emph{European Conference on Computer Vision}.\hskip 1em plus 0.5em minus 0.4em\relax Springer, 2024, pp. 363--380.

\bibitem{qin2019infrared}
Y.~Qin, L.~Bruzzone, C.~Gao, and B.~Li, ``Infrared small target detection based on facet kernel and random walker,'' \emph{IEEE Transactions on Geoscience and Remote Sensing}, vol.~57, no.~9, pp. 7104--7118, 2019.

\bibitem{han2020infrared}
J.~Han, S.~Moradi, I.~Faramarzi, H.~Zhang, Q.~Zhao, X.~Zhang, and N.~Li, ``Infrared small target detection based on the weighted strengthened local contrast measure,'' \emph{IEEE Geoscience and Remote Sensing Letters}, vol.~18, no.~9, pp. 1670--1674, 2020.

\bibitem{zhang2018road}
Z.~Zhang, Q.~Liu, and Y.~Wang, ``Road extraction by deep residual u-net,'' \emph{IEEE Geoscience and Remote Sensing Letters}, vol.~15, no.~5, pp. 749--753, 2018.

\bibitem{sun2023receptive}
H.~Sun, J.~Bai, F.~Yang, and X.~Bai, ``Receptive-field and direction induced attention network for infrared dim small target detection with a large-scale dataset irdst,'' \emph{IEEE Transactions on Geoscience and Remote Sensing}, vol.~61, pp. 1--13, 2023.

\bibitem{hou2021ristdnet}
Q.~Hou, Z.~Wang, F.~Tan, Y.~Zhao, H.~Zheng, and W.~Zhang, ``Ristdnet: Robust infrared small target detection network,'' \emph{IEEE Geoscience and Remote Sensing Letters}, vol.~19, pp. 1--5, 2021.

\bibitem{han2019local}
J.~Han, S.~Moradi, I.~Faramarzi, C.~Liu, H.~Zhang, and Q.~Zhao, ``A local contrast method for infrared small-target detection utilizing a tri-layer window,'' \emph{IEEE Geoscience and Remote Sensing Letters}, vol.~17, no.~10, pp. 1822--1826, 2019.

\bibitem{rosenfeld1976digital}
A.~Rosenfeld, \emph{Digital picture processing}.\hskip 1em plus 0.5em minus 0.4em\relax Academic press, 1976.

\bibitem{unser1995texture}
M.~Unser, ``Texture classification and segmentation using wavelet frames,'' \emph{IEEE Transactions on image processing}, vol.~4, no.~11, pp. 1549--1560, 1995.

\bibitem{do2002wavelet}
M.~N. Do and M.~Vetterli, ``Wavelet-based texture retrieval using generalized gaussian density and kullback-leibler distance,'' \emph{IEEE transactions on image processing}, vol.~11, no.~2, pp. 146--158, 2002.

\bibitem{cohen1992biorthogonal}
A.~Cohen, I.~Daubechies, and J.-C. Feauveau, ``Biorthogonal bases of compactly supported wavelets,'' \emph{Communications on pure and applied mathematics}, vol.~45, no.~5, pp. 485--560, 1992.

\bibitem{shensa2002discrete}
M.~J. Shensa, ``The discrete wavelet transform: wedding the a trous and mallat algorithms,'' \emph{IEEE Transactions on signal processing}, vol.~40, no.~10, pp. 2464--2482, 2002.

\end{thebibliography}

\end{document}